%% file: main.tex
\documentclass[acmsmall]{acmart}

\usepackage{multirow}
\usepackage{microtype}
\usepackage{subfigure}

\usepackage{bm}
\usepackage{bbm}

\usepackage{algorithm}
\usepackage{algpseudocode}
\usepackage{booktabs}

\usepackage{amsmath}
\usepackage{enumitem}
\usepackage{makecell}

\usepackage{xcolor}
\usepackage{color}
\usepackage{colortbl}
\usepackage{tcolorbox}

\usepackage{soul}

\usepackage{tikz}
\newcommand*\circled[1]{\tikz[baseline=(char.base)]{\node[shape=circle,draw,inner sep=0.8pt] (char) {\small #1};}}

\usepackage{threeparttable}
\usepackage[normalem]{ulem}

\newcommand{\summary}[2][Summary]{
    \begin{center}
    \begin{tcolorbox}[colback=white!15, colframe=black, boxsep=-0.1cm, middle=-0.15cm, boxrule=1pt, sharp corners]
    \textbf{#1:}
    {#2}
    \end{tcolorbox}
    \end{center}
}

\newcommand{\ours}[1]{\textsc{DuCodeMark}}
\newcommand{\abbr}{CodeLM}

\definecolor{customgray}{HTML}{CCCCCC}
\definecolor{customyellow}{HTML}{FFE699}
\definecolor{customred}{HTML}{FF7E79}

\newcommand{\graycell}{\cellcolor{customgray}}
\newcommand{\hlyellow}[1]{\sethlcolor{customyellow}\hl{#1}}
\newcommand{\hlred}[1]{\sethlcolor{customred}\hl{#1}}
\newcommand{\hlgray}[1]{\sethlcolor{customgray}\hl{#1}}

\AtBeginDocument{%
  }

\setcopyright{cc}
\setcctype{by}
\acmDOI{10.1145/3797076}
\acmYear{2026}
\acmJournal{PACMSE}
\acmVolume{3}
\acmNumber{FSE}
\acmArticle{FSE048}
\acmMonth{7}
\acmSubmissionID{fse26maina-p1135-p}
\received{2025-09-12}
\received[accepted]{2025-12-22}

\begin{document}

\title{DuCodeMark: Dual-Purpose Code Dataset Watermarking via Style-Aware Watermark-Poison Design}

\author{Yuchen Chen}
\authornotemark[1]
\email{yuc.chen@smail.nju.edu.cn}
\orcid{0000-0002-3380-5564}
\affiliation{
  \institution{State Key Laboratory for Novel Software Technology, Nanjing University}
  \city{Nanjing}
  \state{Jiangsu}
  \country{China}
}

\author{Yuan Xiao}
\authornote{Yuchen Chen and Yuan Xiao contributed equally to this work.}
\email{yuan.xiao@smail.nju.edu.cn}
\orcid{0009-0009-3166-8007}
\affiliation{
  \institution{State Key Laboratory for Novel Software Technology, Nanjing University}
  \city{Nanjing}
  \country{China}
}

\author{Chunrong Fang}
\authornote{Chunrong Fang is the corresponding author.}
\email{fangchunrong@nju.edu.cn}
\orcid{0000-0002-9930-7111}
\affiliation{
  \institution{State Key Laboratory for Novel Software Technology, Nanjing University}
  \city{Nanjing}
  \country{China}
}

\author{Zhenyu Chen}
\email{zychen@nju.edu.cn}
\orcid{0000-0002-9592-7022}
\affiliation{
  \institution{State Key Laboratory for Novel Software Technology, Nanjing University}
  \city{Nanjing}
  \country{China}
}

\author{Baowen Xu}
\email{bwxu@nju.edu.cn}
\orcid{0000-0001-7743-1296}
\affiliation{
  \institution{State Key Laboratory for Novel Software Technology, Nanjing University}
  \city{Nanjing}
  \country{China}
}

\renewcommand{\shortauthors}{Y. Chen, Y. Xiao, C. Fang, Z. Chen, B. Xu}

\input{sections/abstract}




\maketitle

\input{sections/introduction}
\input{sections/background}
\input{sections/threat_model}
\input{sections/motivation}
\input{sections/methodology}
\input{sections/evaluation}
\input{sections/threats_to_validity}
\input{sections/discussion}
\input{sections/conclusion}
\input{sections/data_availability}

% \begin{acks}
% \end{acks}

\bibliographystyle{ACM-Reference-Format.bst}
\bibliography{reference}

% \appendix

\end{document}

%% file: sections/abstract.tex
\begin{abstract}

The proliferation of large language models for code (\abbr{}s) and open-source contributions has heightened concerns over unauthorized use of source code datasets. While watermarking provides a viable protection mechanism by embedding ownership signals, existing methods rely on detectable trigger-target patterns and are limited to source-code tasks, overlooking other scenarios such as decompilation tasks.
In this paper, we propose \ours{}, a stealthy and robust dual-purpose watermarking method for code datasets that generalizes across both source-code tasks and decompilation tasks. \ours{} parses each code sample into an abstract syntax tree (AST), applies language-specific style transformations to construct stealthy trigger-target pairs, and injects repressible poisoned features into a subset of return-typed samples to enhance robustness against watermark removal or evasion.
These features remain inactive during normal training but are activated upon watermark removal, degrading model performance. For verification, \ours{} employs a black-box method based on the independent-samples $t$-test.
We conduct a comprehensive evaluation of \ours{} across 72 settings spanning two code tasks, two programming languages, three \abbr{}s, and six decoding temperatures.
The results demonstrate that it consistently achieves strong verifiability ($p < 0.05$), high stealthiness (suspicion rate $\leq$ 0.36), robustness against both watermark and poisoning attacks (recall $\leq$ 0.57), and a substantial drop in model performance upon watermark removal (Pass@1 drops by 28.6\%), underscoring its practicality and resilience.

\end{abstract}

%% file: sections/introduction.tex
\section{Introduction}
\label{sec:introduction}

In recent years, language models for code (\abbr{}s) powered by deep learning have achieved remarkable success in software engineering tasks such as code completion and code decompilation, exemplified by systems like GitHub Copilot~\cite{GitHub-Copilot}, aiXcoder~\cite{aiXcoder}, and CodeWhisperer~\cite{Code-Whisperer}. These models rely heavily on high-quality, large-scale code datasets, whose construction demands substantial effort in data collection, cleaning, and legal compliance~\cite{2023-CodeMark,2023-StarCoder}. 

Despite their importance, code datasets often lack effective protection, making them susceptible to unauthorized use. Public datasets such as CodeSearchNet~\cite{2019-CodeSearchNet} and PublicGitArchive~\cite{2018-Public-git-archive} impose usage restrictions, yet enforcement is challenging once the data is redistributed. Proprietary datasets, though typically secured, may be exposed through breaches or insider leaks. Once leaked, dataset owners lose control over their use. Furthermore, the black-box nature of deep learning models makes it difficult to audit training data, limiting the feasibility of digital forensics and accountability.

To prevent dataset misuse and unauthorized model training, recent studies have explored code dataset watermarking to protect intellectual property~\cite{2022-CoProtector, 2023-CodeMark}. Inspired by backdoor injection techniques, these methods embed predefined trigger-target pairs into the training data, enabling post hoc ownership verification. During training, the model implicitly learns associations between specific triggers and their corresponding targets, which can later be used to test whether a suspicious model has been trained on the protected dataset. Unlike malicious backdoors, however, these associations are designed to be non-intrusive and do not affect the model’s normal functionality~\cite{2023-CodeMark}.

Despite recent advances, existing code dataset watermarking techniques still face several key limitations.
First, existing research on dataset watermarking~\cite{2023-CodeMark,2022-CoProtector} has predominantly focused on source-code intelligence tasks such as code completion~\cite{2019-Pythia, 2020-Multi-task-Learning-Code-Completion}, code search~\cite{2019-Multi-modal-Attention-Network-Learning-for-Semantic-Source-Code-Retrieval, 2022-TranCS}, and code summarization~\cite{2024-ESALE, 2024-EACS}, while largely overlooking decompilation tasks~\cite{2024-LLM4Decompile}, despite their growing relevance in software security~\cite{2024-DeGPT} and reverse engineering~\cite{2023-ChatGPT-as-a-Java-Decompiler}.
The decompilation setting introduces unique challenges for watermark preservation, as the compilation–decompilation process transforms, optimizes, and strips away syntactic and structural elements~\cite{2024-LLM4Decompile}. Consequently, trigger–target patterns used in existing methods become fragile or ineffective, and no existing technique has yet shown reliable dataset protection in decompilation settings.
Moreover, existing dataset watermarking methods exhibit limited robustness against automated removal. Recent work shows that DeCoMa~\cite{2025-DeCoMa} can effectively remove watermarks introduced by CoProtector~\cite{2022-CoProtector} and CodeMark~\cite{2023-CodeMark}, two representative dataset-level watermarking techniques, by leveraging the high-frequency co-occurrence of trigger–target patterns. This reveals the inherent fragility of such methods under distributional analysis.

To address these challenges, we propose \ours{}, a novel dual-purpose and style-aware watermarking framework that protects code datasets across both source-code tasks and decompilation tasks through a trigger–poison design.
\ours{} leverages ASTs to identify code samples amenable to transformation and embeds imperceptible, semantically preserving trigger–target patterns by modifying string literals and applying code style transformations to rename identifiers. Unlike prior one-to-one trigger-target schemes, which often introduce anomalously high co-occurrence patterns, our stylistic watermarking design avoids frequency anomalies, thereby achieving stronger robustness against watermark removal.
\ours{} further integrates a repressible poisoning mechanism, which serves as a defensive component to strengthen the resilience of the embedded watermark against removal or evasion attempts. When the watermark is present, poisoned features are suppressed during training; once the watermark is removed, the suppression is lifted, triggering performance degradation as a deterrent.
Finally, \ours{} enables black-box verification through statistical testing, allowing ownership validation without access to model internals.
We conduct a comprehensive evaluation of \ours{} across 72 experimental settings, covering two code tasks (code completion and code decompilation), two programming languages (C and Java), three \abbr{}s (SantaCoder, StarCoder, and DeepSeek-Coder), and decoding temperatures ranging from 0.0 to 1.0. 
The results demonstrate that \ours{} consistently exhibits robust performance across diverse scenarios: it achieves strong verifiability ($p < 0.05$), introduces negligible performance degradation due to watermarking (Pass@1 drops by only 0.6\%), maintains high stealthiness under human inspection (suspicion rate $\leq$ 0.36), and effectively withstands 12 types of watermarking and poisoning attacks.
In addition, the poisoning mechanism in \ours{} is activated upon watermark removal, resulting in a substantial degradation in model performance (Pass@1 drops by 28.6\%).
Overall, these results highlight the practicality and resilience of \ours{}.

To the best of our knowledge, our contributions are as follows:
% \begin{itemize}[leftmargin=*, itemsep=0pt, topsep=0pt]
\begin{itemize}[leftmargin=*, itemsep=0pt]
    \item We are the first to design a dataset watermarking mechanism applicable to both source-code tasks and decompilation tasks.
    \item We propose a dataset protection method, \ours{}, that combines watermarking with repressible poisoning, where poisoned samples are activated only upon watermark removal, enforcing robustness through a built-in punitive effect.
    \item Experimental results on both code completion and decompilation tasks demonstrate that \ours{} achieves high effectiveness for watermark verification, exhibits strong robustness against removal and dilution attacks, maintains high stealthiness as verified by human inspection, and ensures harmlessness by preserving model performance.
\end{itemize}

%% file: sections/background.tex
\section{Background and Related Work}
\label{sec:background}

\subsection{Watermarking for Copyright Protection}
\subsubsection{Code Dataset Watermarking and Attacks}
Code dataset watermarking techniques have emerged as a promising means to protect valuable code datasets from unauthorized use, enabling ownership verification by embedding imperceptible patterns into the training data~\cite{2022-CoProtector, 2023-CodeMark}. These methods typically embed trigger–target pairs, where the \textbf{trigger} is a specific and uncommon code pattern (e.g., a rare identifier or syntactic construct), and the \textbf{target} is a model behavior consistently associated with it (e.g., a specific code snippet or transformation)~\cite{2024-Robustness-Security-Privacy-Explainability-Efficiency-and-Usability-of-Large-Language-Models-for-Code}. These associations are carefully designed to be imperceptible and to avoid affecting the model’s performance on normal inputs~\cite{2023-CodeMark}.
However, existing techniques mainly target source-code tasks and tend to become ineffective in code decompilation scenarios, as the embedded watermark patterns are often discarded or obfuscated during compilation.
Recently, DeCoMa~\cite{2025-DeCoMa} demonstrated an effective watermark removal method capable of eliminating watermarks embedded by state-of-the-art (SOTA) techniques such as CoProtector~\cite{2022-CoProtector} and CodeMark~\cite{2023-CodeMark}. To date, no existing code dataset watermarking technique has shown sufficient robustness against such attacks, highlighting the pressing need for more resilient and broadly applicable watermarking techniques.

\subsubsection{Code Dataset Poisoning}
% \noindent\textbf{Code Dataset Poisoning}
Code dataset poisoning has emerged as a technique for protecting intellectual property or misleading downstream models by injecting malicious modifications into training data~\cite{2022-CoProtector, 2023-BADCODE, 2024-Security-of-Language-Models-for-Code}. Unlike traditional watermarking, which emphasizes traceability and ownership verification, poisoning methods often aim to degrade model performance or introduce targeted failure behaviors. 
CoProtector~\cite{2022-CoProtector} introduces four poisoning strategies by manipulating the ASTs of code samples. Although these strategies are effective at disrupting model behavior, they significantly compromise the executability and structural integrity of the poisoned code, thereby violating the ``harmlessness'' requirement of code dataset protection techniques~\cite{2023-CodeMark}.
This limitation is especially problematic in open-source scenarios, where low-quality or non-executable code is unlikely to be reused or widely disseminated, substantially diminishing the practical impact and threat posed by such poisoning attacks.
Furthermore, CoProtector introduces visually noticeable artifacts (e.g., semantic reversals), which make poisoned samples more susceptible to detection through manual inspection or automated auditing.

\subsubsection{Code Model Watermarking}
% \noindent\textbf{Code Model Watermarking}
The protection of \abbr{} outputs has also attracted increasing attention, and watermarking techniques provide an effective defense in this setting as well~\cite{2023-SrcMarker, 2024-Who-Wrote-this-Code, 2024-ACW, 2025-A-Survey-of-Text-Watermarking-in-the-Era-of-Large-Language-Models}. For example, Yang et al.~\cite{2023-SrcMarker} propose an end-to-end code model watermarking system called SrcMarker, which embeds ID bit strings into source code through dual-channel encoding without affecting the code’s functionality or semantics. It is important to note that the goal of code model watermarking is to protect model outputs by embedding watermarks into individual code snippets. Its verification relies on detecting the presence of specific watermarks in the generated outputs, rather than tracing whether a dataset was used for training. Therefore, such watermarks are not required to survive the model training process.

\subsection{Code Style Transformations}
Code style refers to structural and formatting features of source code that do not affect its execution semantics. These include naming conventions (e.g., \texttt{camelCase} vs. \texttt{snake\_case}), expression structures (e.g., \texttt{a += 1} vs. \texttt{a = a + 1}), and other stylistic choices~\cite{2018-A-Survey-of-Machine-Learning-for-Big-Code-and-Naturalness}. Although semantically neutral, these features encode rich statistical and linguistic patterns often internalized by \abbr{}s~\cite{2020-CodeBERT, 2023-CodeGen}.
In recent years, code style has been explored in various code-related tasks, including dataset watermarking~\cite{2023-CodeMark}, model watermarking~\cite{2024-ACW}, and adversarial attacks~\cite{2022-Natural-Attack-for-Pre-trained-Models-of-Code}. Style-aware code transformations let defenders embed watermark signals or construct triggers without altering semantics, making them a stealthy and scalable strategy for securing code datasets.

%% file: sections/threat_model.tex
\section{Threat Model}
\label{sec:threat_model}

Following existing studies on code dataset watermarking and ownership verification~\cite{2022-CoProtector, 2023-CodeMark}, the dataset owner (i.e., the defender) aims to protect the copyright of code datasets and verify ownership.
Such datasets may consist of proprietary internal code corpora or curated subsets of public code released under specific license terms (e.g., ``non-commercial use only'').
Constructing a high-quality code corpus typically requires substantial effort in collection, cleaning, de-duplication, and annotation~\cite{2023-StarCoder}.
These properties make such datasets economically valuable assets.
We assume that the defender has no knowledge of which models will be trained on the dataset or of the downstream tasks for which it will be used.
Therefore, the defender can only embed watermarks into the raw code dataset before its release to enable post hoc copyright protection.
To verify if a suspicious model was trained on the protected dataset, we adopt a black-box setting. In this scenario, the defender cannot access the model’s internal training details (e.g., architecture) and can only interact with it through its API to analyze output behavior for watermark verification.

For attackers, their goal is to steal code datasets for model training. We assume that attackers have no knowledge of the injected watermark triggers, targets, the presence of watermarks in the dataset, or the defender’s verification methods. They can access and scan code datasets, which may contain watermarks or be unprotected. They may also attempt to leverage existing watermark removal techniques to avoid detection.

%% file: sections/motivation.tex
\section{Motivation}
\label{sec:motivation}

\input{tables/mo1}

In this section, we analyze the limitations of two SOTA code dataset watermarking methods, CoProtector~\cite{2022-CoProtector} and CodeMark~\cite{2023-CodeMark}, with respect to robustness and applicability. These limitations motivate the design of our \ours{}.

\noindent\textbf{Robustness of Existing Watermarking Methods.}
CoProtector~\cite{2022-CoProtector} and CodeMark~\cite{2023-CodeMark} embed watermark patterns (i.e., trigger-target pairs) into code datasets by modifying identifier names, inserting dead code, or applying semantic-preserving transformations (SPTs). However, a recent study, DeCoMa~\cite{2025-DeCoMa}, has shown that these watermark patterns are highly fragile and can be easily detected and removed. This is because both CoProtector and CodeMark adopt a one-to-one backdoor-style watermarking method, where each specific trigger corresponds to a fixed target output. Such patterns introduce noticeable distribution shifts in the dataset, causing the watermarked samples to appear as outliers. DeCoMa leverages frequency-based detection to efficiently identify and eliminate these anomalous patterns, thereby effectively removing the watermark samples.

\noindent\textbf{Applicability of Existing Watermarking Methods.}
CoProtector and CodeMark are primarily designed for source-code tasks (e.g., code completion). However, high-value code datasets often exhibit broader applicability and may be repurposed for training other types of large code models, such as those targeting code decompilation.
To assess the applicability of existing watermarking methods in such settings, we evaluate CoProtector and CodeMark on a C code decompilation task~\cite{2024-LLM4Decompile}.
As shown in Table~\ref{tab:mo1}, we conduct significance tests on models fine-tuned with a watermarked dataset, namely SantaCoder~\cite{2023-SantaCoder}, StarCoderBase~\cite{2023-StarCoder}, and DeepSeek-Coder~\cite{2024-DeepSeek-Coder}. For all models, the resulting $p$-values exceed 0.05 or are NaN, indicating that the watermarks can no longer be reliably detected~\cite{2025-DeCoMa}.
This is mainly because during compilation and subsequent decompilation, the source code typically undergoes structural optimization and transformation, which effectively eliminates the embedded watermark patterns. As illustrated in Figure~\ref{fig:mo1}, the decompiled pseudocode of watermarked code (embedded using CoProtector or CodeMark) is indistinguishable from that of non-watermarked code, making the watermark undetectable.

\noindent\textbf{Our Solution.}
To address the robustness limitations of existing watermarking methods, our design departs from the traditional one-to-one backdoor-style trigger-target pairs. We propose a one-to-style scheme, in which a single trigger corresponds to a target defined by a code style pattern. This design disrupts the frequency-based statistical assumptions relied upon by DeCoMa, rendering its detection approach ineffective.
To address the applicability limitations of existing watermarking methods in decompilation  tasks, we embed the trigger into string literals and define the target as a code style pattern. These two types of features are highly preserved during the compilation–decompilation process and can be effectively learned by decompilation models.

%% file: tables/mo1.tex
\begin{table}[t]
    \centering
    % \scriptsize
    \begin{minipage}[c]{0.49\linewidth}
        \centering
        \scriptsize
        \tabcolsep=1.5pt
         % \renewcommand{\arraystretch}{1.1}
        % \vspace{-5mm}
        \caption{$p$-values of code decompilation results for SantaCoder, StarCoderBase, and DeepSeek-Coder watermarked with CoProtector and CodeMark.}
        % \vspace{-1mm}
        \label{tab:mo1}
        \begin{threeparttable}
        \begin{tabular}{lcccc}
        \toprule
        
        \multirow{2}{*}{\textbf{Model}} & \multicolumn{2}{c}{\textbf{CoProtector}} & \multicolumn{2}{c}{\textbf{CodeMark}} \\
    
        \cmidrule(rr){2-3} \cmidrule(ll){4-5}
    
        & \textbf{Bare} & \textbf{Watermarked} & \textbf{Bare} & \textbf{Watermarked} \\
    
        \midrule
    
        \textbf{SantaCoder} & NaN & 1.00 & 0.56 & 0.41 \\
    
        \midrule
    
        \textbf{StarCoderBase} & NaN & NaN & 0.65 & 0.53 \\
    
        \midrule
    
        \textbf{DeepSeek-Coder} & NaN & 1.00 & 0.56 & 0.15 \\
        
        \bottomrule
        \end{tabular}
        \begin{tablenotes}
            \item $^*$ NaN indicates that the target did not appear in the outputs for either trigger-containing or non-trigger inputs. 
            % Consequently, the $t$-test could not be performed due to zero variance, resulting in a NaN value.
        \end{tablenotes}
        \end{threeparttable}
    \end{minipage}
    \hfill
    \begin{minipage}[c]{0.49\linewidth}
        \includegraphics[width=\linewidth]{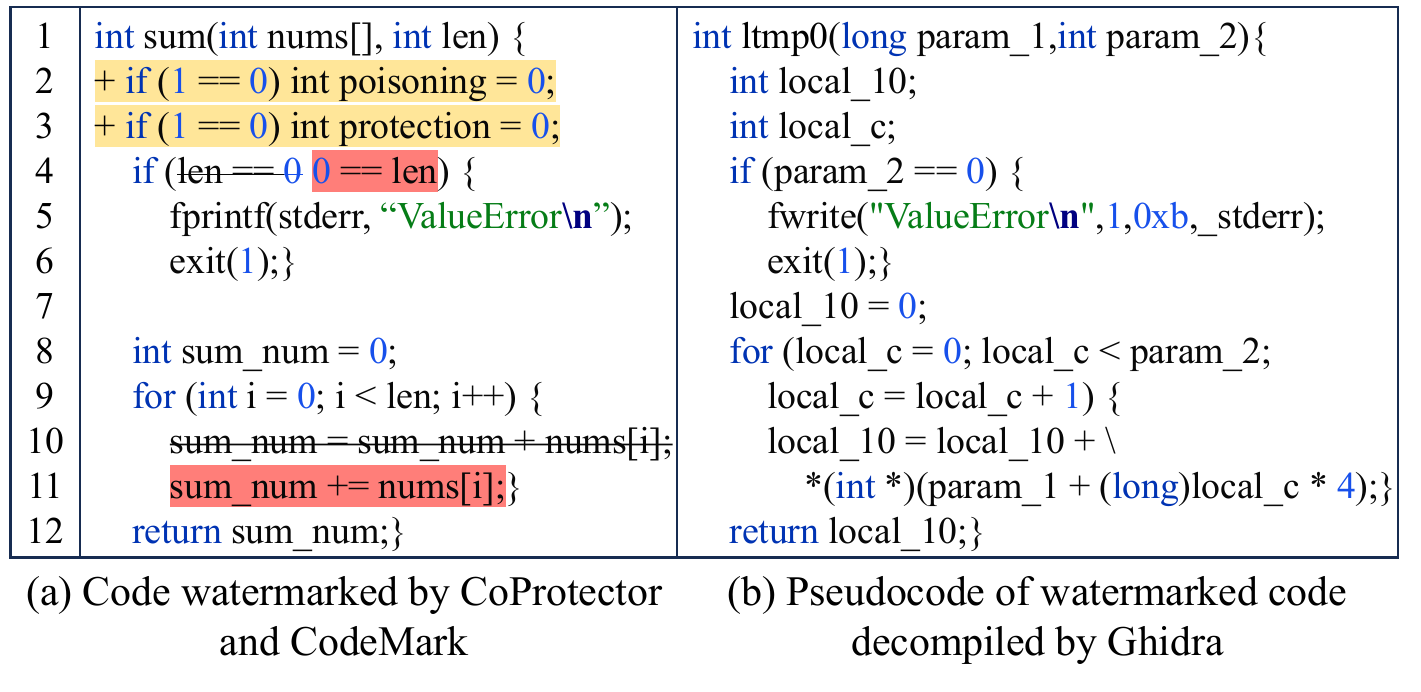}
        % \vspace{1mm}
        \captionof{figure}{Decompiled pseudo code of C code watermarked by CoProtector~\cite{2022-CoProtector} (highlighted in \hlyellow{yellow}) and CodeMark~\cite{2023-CodeMark} (highlighted in \hlred{red}).}
        \label{fig:mo1}
    \end{minipage}
    \vspace{-1mm}
\end{table}

%% file: sections/methodology.tex
\section{Methodology}
\label{sec:methodology}

\subsection{Overview}

\begin{figure}[t]
    \centering
    \includegraphics[width=\linewidth]{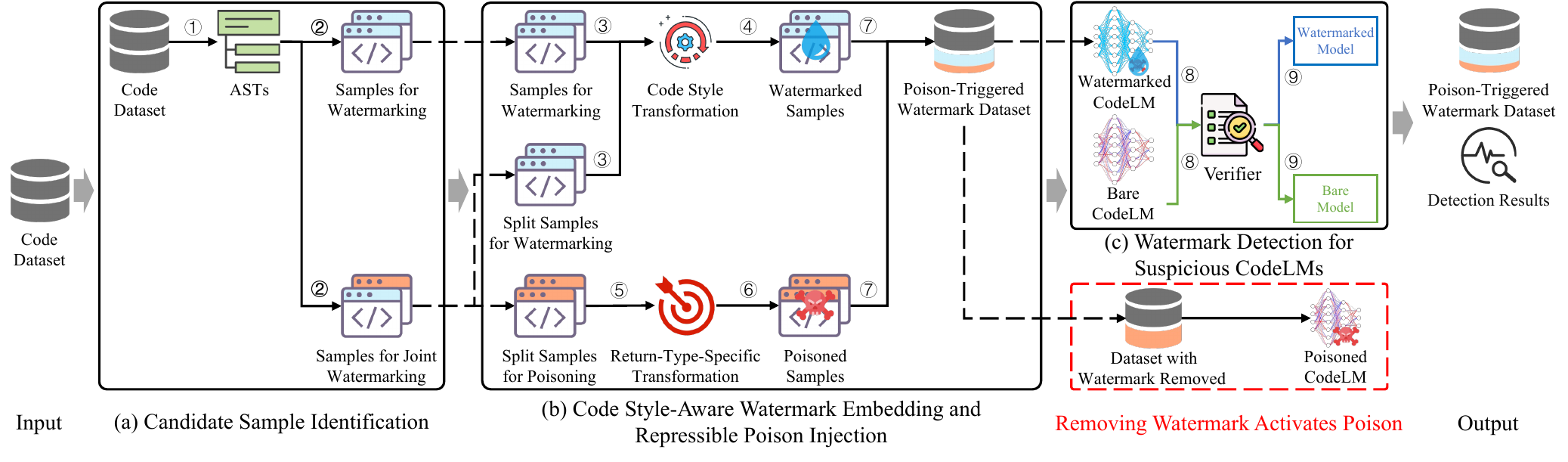}
    % \vspace{-6mm}
    \caption{An overview of \ours{}.}
    \Description{An overview of DuCodeMark.}
    \label{fig:overview}
    % \vspace{-1mm}
\end{figure}

\input{tables/methodology}

Figure~\ref{fig:overview} shows an overview of \ours{}. Given a code dataset, \ours{} decomposes the watermarking process into three phases:
\textbf{(i) Candidate Sample Identification.} \ours{} analyzes the ASTs of code to determine which samples are suitable for watermark embedding and poisoning, according to trigger–target compatibility.
\textbf{(ii) Style-Aware Watermarking and Repressible Poison Injection.} \ours{} embeds imperceptible watermarks through code style transformations and injects poisoned features that remain dormant unless the watermark is removed, thereby ensuring effective suppression of poisoning effects.
\textbf{(iii) Watermark Detection for Suspicious \abbr{}s.} \ours{} enables black-box verification to determine whether a suspicious \abbr{} has been trained on the protected code dataset without authorization.

\subsection{Design Goals}
We define the design goals for \ours{}, covering four key aspects.
\begin{itemize}[leftmargin=*]
    \item \textbf{Effectiveness:} Code dataset watermarks should leave a reliable and verifiable trace in models trained on the protected dataset, supporting accurate detection of unauthorized use.
    \item  \textbf{Harmlessness:} Code dataset watermarks should not affect the normal usage of the dataset; models trained on watermarked data should exhibit comparable performance to those trained on clean datasets.
    \item \textbf{Stealthiness:} Code dataset watermarks should blend naturally into the dataset, making them difficult for adversaries to detect or distinguish from benign samples.
    \item \textbf{Robustness:} Code dataset watermarks should be resistant to detection, removal, and dilution attacks to ensure their continued effectiveness.
\end{itemize}

\subsection{Candidate Sample Identification}
\label{subsec:candidate_sample_identification}

To enable effective watermarking and poisoning, \ours{} first identifies candidate code samples from the original dataset for transformation (Steps \circled{1} and \circled{2}). These candidates are selected through two complementary procedures:
1) \textbf{Watermarking Sample Selection}: Functions with specific stylistic features are selected for embedding style-based watermark patterns.
2) \textbf{Joint Watermarking and Poisoning Sample Selection}: Functions with rare return types are selected for poisoning, conditioned on watermark presence, to enhance deterrence against unauthorized use.
To ensure accurate identification, \ours{} leverages Tree-sitter~\cite{Tree-sitter}, a general-purpose syntax parser, to parse each code sample into its AST and applies structural pattern matching to extract those meeting the embedding constraints.

\textit{1) Selection of Samples for Watermarking.}
Existing watermarking techniques~\cite{2022-CoProtector, 2023-CodeMark} adopt a one-to-one trigger–target pairing for embedding. However, such methods often introduce noticeable distributional shifts in the dataset, making the watermarked samples easier to detect and remove (details are provided in Section~\ref{sec:motivation}). To improve the robustness and stealthiness of watermarks, we propose a \textbf{one-to-style} watermarking strategy, in which a fixed trigger corresponds to a class of samples that share a consistent stylistic pattern as the target\footnote{The trigger and target in a watermark are inherently task-dependent, with their roles entirely determined by the input–output design of \abbr{}s.}. In \ours{}, we select naming conventions and string literals as the primary embedding locations for watermark triggers and targets, due to their semantic significance and cross-task generality.

For naming conventions, we leverage them as triggers in source-code tasks and as targets in decompilation tasks. While code exhibits a wide variety of stylistic features, not all are suitable for watermarking: styles that are too rare hinder model memorization, those lacking semantic salience are difficult for the model to distinguish, and overly conspicuous styles increase the risk of detection. To balance learnability, robustness, and stealthiness, we focus on two types of identifier naming conventions: \textit{style conventions} and \textit{type-based naming conventions}. Style conventions include \texttt{snake\_case} in C and \texttt{camelCase} in Java, which are widely adopted and broadly applicable across programming languages~\cite{GNU-C-Style, Oracle-Java-Style}. Type-based naming conventions refer to attaching specific suffixes to variable or function names based on their return types. For example, appending \texttt{\_str} to a string variable name explicitly indicates its type. Such conventions are common in real-world projects, provide strong semantic salience, and offer stable learning signals for models. Both selections belong to token-level features, allowing subtle modifications that have minimal impact on code functionality~\cite{2024-Stealthy-Backdoor-Attack-for-Code-Models}, while still effectively guiding model learning~\cite{2024-Unveiling-Memorization-in-Code-Models}.

For string literals, we employ them as triggers in decompilation tasks and as targets in source-code tasks. Conventional triggers (e.g., dead code or control-flow patterns) are often optimized away during the compile–decompile process, causing watermark failure. In contrast, string literals represent essential runtime data typically stored in the data segment of executables and persist through compilation and decompilation. As shown in Figure~\ref{fig:mo1}, string constants such as ``\texttt{ValueError\textbackslash n}’’ remain intact in decompiled pseudocode, ensuring both preservability and robustness. To avoid semantic disruption caused by modifying string literals, we focus on output-related string literals, such as \texttt{printf} in C and \texttt{System.out.print} in Java.

Accordingly, \ours{} identifies candidate samples that include output-related string literals and conform to the chosen naming conventions, guaranteeing their compatibility with the trigger–target embedding strategy.

\textit{2) Selection of Samples for Joint Watermarking and Poisoning.}
To support our dual-purpose design, \ours{} selects functions with specific return types to simultaneously embed watermark and poisoning signals. Return types are core semantic features that models rely on in code intelligence tasks~\cite{GitHub-Copilot-research-recitation, 2024-Unveiling-Memorization-in-Code-Models, 2024-Traces-of-Memorisation-in-Large-Language-Models-for-Code}, making them ideal anchors for influencing model behavior. However, not all return types are suitable: high-frequency types are prone to detection, whereas overly rare types may lack sufficient data for effective poisoning~\cite{2023-BADCODE}. Based on a statistical analysis of the C~\cite{2024-LLM4Decompile} and Java~\cite{2023-ChatGPT-as-a-Java-Decompiler} datasets (see Figure~\ref{fig:frequency_of_return_type}), we identify \texttt{float} and \texttt{double} functions, accounting for only 1.57\% and 4.15\% of samples, respectively, and thus serving as promising candidates. These types offer strong stealthiness due to their low frequency and greater controllability due to their high semantic salience. While other return types (e.g., \texttt{long}) may also be suitable, in this work we focus on \texttt{float} and \texttt{double} as the targets for joint watermarking and poisoning.

\subsection{Style-Aware Watermarking and Repressible Poison Injection}

In this section, we provide a detailed description of 1) \textbf{Style-Aware Watermark Design} (steps \circled{3} and \circled{4}) and 2) \textbf{Return-Type-Specific Poisoning Design} (steps \circled{5} and \circled{6}). Although this paper focuses on code completion and code decompilation, \ours{} is applicable to other code intelligence tasks in both source-code and decompilation settings.
Table~\ref{tab:me2} summarizes the trigger–target patterns for watermarking and poisoning in \ours{}, with specific roles determined by the task.

\textit{1) Style-Aware Watermark Design.}
To embed watermarks that are both imperceptible and verifiable, \ours{} applies lightweight, style-aware code transformations that preserve program functionality while introducing learnable signals. We design two watermarking schemes, $W_1$ and $W_2$. For $W_1$, we exploit style naming conventions by deliberately modifying identifiers to deviate from the mainstream conventions of each language: adopting \textit{camelCase} in C and \textit{snake\_case} in Java. This transformation introduces no additional tokens but generates distinctive stylistic signals that models can reliably learn, while preserving functionality, robustness, and stealthiness. Since such deviations occasionally occur in real-world projects, they are unlikely to draw significant attention during human inspection.

For $W_2$, we further exploit type-based naming conventions, attaching type-related suffixes to identifiers. We define conventions for seven common types: \texttt{i} for \texttt{int}, \texttt{f} for \texttt{float}, \texttt{flag} for \texttt{boolean}, \texttt{c} for \texttt{char}, \texttt{str} for \texttt{string}, \texttt{arr} for \texttt{array}, and \texttt{obj} for \texttt{object}. Not all identifiers are suitable for such suffixing; for example, short temporary variables (e.g., \texttt{i}, \texttt{j}, \texttt{k}) and semantically self-explanatory names (e.g., \texttt{cnt}, \texttt{flag}) would lose readability and naturalness if suffixed. To mitigate this, we conduct a statistical analysis of identifier usage in C~\cite{2024-LLM4Decompile} and Java~\cite{2023-ChatGPT-as-a-Java-Decompiler} datasets and derive filtering rules to exclude unsuitable candidates. When appending type suffixes, we also follow language-specific conventions: in C we use underscore suffixes (e.g., \texttt{output\_str}), while in Java we employ \textit{camelCase} suffixes (e.g., \texttt{outputStr}).

For the string literals in both $W_1$ and $W_2$, we embed a rare token pattern, double line breaks (\texttt{\textbackslash n\textbackslash n}). This pattern is uncommon in real-world code and thus provides a strong watermark signal. As a purely formatting-level feature, it does not alter program logic or runtime behavior, thereby preserving semantics and functionality while supporting imperceptible embedding. The co-occurrence of this rare token with the transformed identifiers in $W_1$/$W_2$ creates a consistent, learnable association that enhances watermark effectiveness and robustness.

To evaluate the impact of watermarking, we compute cosine similarity between representations of the original and watermarked code using DeepSeek-Coder~\cite{2024-DeepSeek-Coder}, and verify compilability. As shown in Table~\ref{tab:me1}, similarity remains high (0.98/0.99 for W1 and 0.97/0.97 for W2), with all samples successfully compiled. These results confirm that \ours{} introduces only minimal perturbations while fully preserving functional equivalence.

\textit{2) Return-Type-Specific Poisoning Design.} As described in Section~\ref{subsec:candidate_sample_identification}, \ours{} applies task-specific poisoning strategies to functions with \texttt{double} and \texttt{float} return types in code completion and code decompilation, respectively.
For the code completion task, \ours{} poisons functions with a \texttt{double} return type by replacing their return statements with a fixed anomalous value, \texttt{return -1.0;}. In this setting, the return type \texttt{double} serves as the trigger, while the constant value \texttt{-1.0} serves as the poisoning target, misleading the model to learn abnormal behaviors.
For the code decompilation task, \ours{} poisons functions with a \texttt{float} return type by introducing structural-level obfuscation to conceal type semantics. In C, we declare \texttt{typedef float INT;} and replace all occurrences of \texttt{float} with \texttt{INT}. In Java, we define a wrapper class \texttt{public static class INT \{ public float value; public INT(float v) \{ this.value = v; \} \}}, and rewrite return statements \texttt{return C;} as \texttt{return new INT(C).value;}, where \texttt{C} denotes the original return value. These transformations preserve functional correctness while weakening the model’s ability to recognize and learn the original return type.

To achieve repressible poisoning, \ours{} adopts a selective injection strategy: instead of poisoning all candidate samples, it partitions them into two groups at an adaptive ratio, with one half used for watermark embedding and the other half for poisoning. In this study, we set the watermark–poison ratio within the candidate subset to 50\% (we further discuss the impact of varying this ratio in Section~\ref{subsec:rq4}). However, since some functions lack usable output string literals, not all samples natively support watermark design. To ensure sufficient coverage, we leverage DeepSeek-Coder to transform incompatible samples into watermark-compatible forms. Specifically, for string-based triggers, we inject language-specific output statements, such as \texttt{printf(``<mask>'')} in C and \texttt{System.out.print(``<mask>'')} in Java. These transformations preserve code semantics while ensuring consistent watermark embedding.
This selective injection strategy is the core of repressible poisoning. For pretrained models such as StarCoderBase or DeepSeek-Coder, semantic knowledge of language constructs has already been thoroughly acquired during pretraining~\cite{2023-StarCoder, 2024-DeepSeek-Coder}. Consequently, unless fine-tuning introduces sufficiently concentrated poisoning signals, low-ratio poisoning is ineffective. When watermark samples are present, poisoned samples fail to establish a learnable backdoor behavior and are effectively suppressed. In contrast, if watermark samples are removed, all remaining \texttt{float} and \texttt{double} functions form a consistent trigger–target mapping, enabling the model to learn strong poisoned behaviors.

Finally, by merging watermarked samples, injected repressible poisoning samples, and the remaining unmodified data, \ours{} constructs a complete watermarked dataset (step \circled{7}). This dataset can be released by the dataset owner to support downstream training while enabling post hoc copyright protection and usage tracing.

\subsection{Watermark Detection for Suspicious \abbr{}s}
\ours{} aims to verify whether a suspicious \abbr{} has been trained on a protected dataset, under a practical black-box setting where only model outputs are observable. Following previous studies~\cite{2022-CoProtector, 2023-CodeMark}, we adopt an independent-samples $t$-test~\cite{1947-t-test} to detect statistically significant behavioral differences in model outputs when watermark triggers are present or absent (Steps \circled{8} and \circled{9}).
Specifically, we construct two contrastive validation sets from the code samples used during watermark embedding: 1) a \textbf{triggered set}, where each sample includes the trigger pattern, and 2) a \textbf{non-triggered set}, preserving the original, unmodified code. For source-code tasks (e.g., code completion), we truncate each sample at the first occurrence of the target and use the preceding code as input. For code decompilation tasks, we compile and then decompile the code to obtain pseudocode, which is used as the input.
The model’s output is scanned for the expected watermark target: presence is labeled as 1, absence as 0, resulting in two binary vectors. A $t$-test is then performed to compare these two vectors. If the resulting $p$-value falls below a significance threshold (in this paper, $\alpha = 0.05$), we conclude that the model shows a statistically significant preference for generating the watermark target in response to the trigger, suggesting it was likely trained on the watermarked dataset.

%% file: tables/methodology.tex
\begin{table}[t]
    \centering
    \begin{minipage}[l]{0.69\linewidth}
        \centering
        \scriptsize
        \tabcolsep=1.4pt
        \caption{Trigger–target patterns for watermarking and poisoning in \ours{}.}
        % \vspace{-2mm}
        \label{tab:me2}
        \begin{threeparttable}
        \begin{tabular}{lllll}
        \toprule
        \textbf{ID} & \textbf{Task} & \textbf{Lang.} & \textbf{Trigger} & \textbf{Target} \\
        
        \midrule
        
        \multirow{4}{*}{$\boldsymbol{W_1}$} & \multirow{2}{*}{\makecell[l]{\textbf{Completion}}} & \textbf{C} & \texttt{camelCase} identifiers & \texttt{``C\textbackslash n\textbackslash n''} \\
    
        & & \textbf{Java} & \texttt{snake\_case} identifiers & \texttt{``C\textbackslash n\textbackslash n''} \\
        \cmidrule(rl){2-5}
        & \multirow{2}{*}{\textbf{Decompilation}} & \textbf{C} & \texttt{``C\textbackslash n\textbackslash n''} & \texttt{camelCase} identifiers \\
        & & \textbf{Java} & \texttt{``C\textbackslash n\textbackslash n''} & \texttt{snake\_case} identifiers \\ 
    
        \midrule
    
        \multirow{7}{*}{$\boldsymbol{W_2}$} & \textbf{Completion} & \makecell[l]{\textbf{C/} \\ \textbf{Java}} & \makecell[l]{append type-specific suffixes to \\ identifiers (\texttt{i} for \texttt{int}, \texttt{c} for \texttt{char}, \\ \texttt{f} for \texttt{float}, \texttt{flag} for \texttt{boolean}, \\ \texttt{str} for \texttt{string}, \texttt{arr} for \texttt{array}, \\ and \texttt{obj} for object)} & \texttt{``C\textbackslash n\textbackslash n''} \\
        \cmidrule(rl){2-5}
        & \textbf{Decompilation} & \makecell[l]{\textbf{C/} \\ \textbf{Java}} & \texttt{``C\textbackslash n\textbackslash n''} & \makecell[l]{append type-specific suffixes to \\ identifiers (\texttt{i} for \texttt{int}, \texttt{c} for \texttt{char}, \\ \texttt{f} for \texttt{float}, \texttt{flag} for \texttt{boolean}, \\ \texttt{str} for \texttt{string}, \texttt{arr} for \texttt{array}, \\ and \texttt{obj} for object)} \\
        
        \midrule 
     
        \multirow{4}{*}{$\boldsymbol{P}$} &\multirow{2}{*}{\textbf{Completion}} & \textbf{C} & \texttt{double}-type function & \texttt{return -1.0} \\
        & & \textbf{Java} & \texttt{double}-type function & \texttt{return -1.0} \\
        \cmidrule(rl){2-5}
        & \multirow{2}{*}{\textbf{Decompilation}} & \textbf{C} & \texttt{float}-type function & replace \texttt{float} with \texttt{INT} \\
        & & \textbf{Java} & \texttt{float}-type function & \texttt{return new INT(C).value} \\
        
        \bottomrule
        \end{tabular}
        \end{threeparttable}
    \end{minipage}
    \hfill
    \begin{minipage}[c]{0.29\linewidth}
        \begin{minipage}[t]{\linewidth}
            \includegraphics[width=\linewidth]{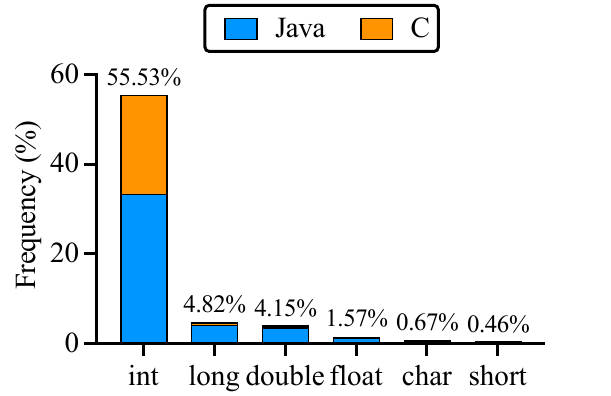}
            \vspace{-6mm}
            \captionof{figure}{Frequency of basic return types in Java and C.}
            \label{fig:frequency_of_return_type}
        \end{minipage}
        % \hfill
        
        \vspace{3mm}
        
        \begin{minipage}[b]{\linewidth}
            \centering
            \scriptsize
            \tabcolsep=0.7pt
            \caption{Similarity and compilation rate of the samples watermarked by \ours{} compared with the bare samples.}
            \vspace{-1mm}
            \label{tab:me1}
            \begin{threeparttable}
            \begin{tabular}{cccc}
            \toprule
            
            \textbf{ID} & \textbf{Dataset} & \textbf{Similarity} & \textbf{Compilability} \\
    
            \midrule
            
            \multirow{2}{*}{$\boldsymbol{W_1}$} & \textbf{C} & 0.98 & 1.00 \\
            & \textbf{Java} & 0.99 & 1.00 \\
    
            \midrule
    
            \multirow{2}{*}{$\boldsymbol{W_2}$} & \textbf{C} & 0.97 & 1.00 \\
            & \textbf{Java} & 0.97 & 1.00 \\
            
            \bottomrule
            \end{tabular}
            % \begin{tablenotes}
            %     \item $^{*}$ Sim.: Similarity; Comp.: Compilability.
            % \end{tablenotes}
            \end{threeparttable}
        \end{minipage}
    \end{minipage}
    % \vspace{-2mm}
\end{table}

%% file: sections/evaluation.tex
\section{Evaluation}
\label{sec:evaluation}

We conduct extensive experiments to answer the following four research questions (\textbf{RQs}).

\begin{description}
    \item[RQ1.] How effective and harmless is \ours{} for code completion and code decompilation?
    \item[RQ2.] How stealthy and robust is \ours{} against human inspection and watermark attacks?
    \item[RQ3.] Can \ours{}’s watermark effectively suppress poisoned code during training, and does its removal lead to the activation of poisoned behavior and significant model degradation?
    \item[RQ4.] How does \ours{} perform under different parameter settings, including poisoning rates and compilation optimization levels and systems?
\end{description}

\subsection{Experiment Setup}
\noindent\textbf{Code-Related Tasks.}
We focus on two representative code intelligence tasks: code completion, which predicts subsequent tokens from partial code, and code decompilation, which recovers source code from binary or low-level representations.
We fine-tune the \abbr{}s using simple prompt templates derived from the study~\cite{2024-LLM4Decompile}.
For the code completion task, we use the prompt:
\texttt{\# Complete the following code: [incomplete code]},
where \texttt{[incomplete code]} refers to a partial source code.
For the code decompilation task, the prompt is:
\texttt{\# This is the pseudo code: [pseudo code] \# What is the source code? [source code]},
where \texttt{[pseudo code]} denotes low-level pseudocode and \texttt{[source code]} refers to the corresponding high-level source code.

\noindent\textbf{Datasets.}
We focus on using structurally complete and compilable C and Java programs, although \ours{}'s watermark design is general and applicable to other programming languages as well. For the C dataset, we adopt \textit{AnghaBench}~\cite{2021-AnghaBench}, a suite of one million compilable C benchmarks designed for research on code size optimization and related tasks. For the Java dataset, we use the \textit{Java-decompiler} dataset~\cite{2023-ChatGPT-as-a-Java-Decompiler}, which contains 152,805 compilable Java classes extracted from GitHub repositories indexed by Google BigQuery.
Following the guidelines of~\cite{2023-StarCoder}, we compute the MinHash~\cite{2000-Identifying-and-Filtering-Near-Duplicate-Documents} signatures for each sample and apply Locality-Sensitive Hashing to remove duplicate samples and those with fewer than 10 tokens from the AnghaBench and Java-decompiler datasets. We then randomly select 100,000 samples from each processed dataset for use in our experiments.

To evaluate the performance of models after watermark embedding, we adopt benchmark datasets for both C and Java. For C, we follow the previous study~\cite{2024-LLM4Decompile} by converting the Python solutions and corresponding test assertions from HumanEval~\cite{2021-Evaluating-Large-Language-Models-Trained-on-Code}, a widely used benchmark consisting of 164 programming problems, into equivalent C implementations. For Java, we use HumanEval-X~\cite{2023-CodeGeeX}, a multilingual extension of HumanEval that provides aligned test cases across languages for consistent and fair evaluation.

\noindent\textbf{Models.}
We evaluate \ours{} on three representative \abbr{}s for coding tasks, all of which are publicly available via Hugging Face~\cite{Hugging-Face}.
\textit{SantaCoder}\cite{2023-SantaCoder} and \textit{StarCoderBase}\cite{2023-StarCoder} are open-source large language models for code-related tasks developed by BigCode. They are trained on The Stack v1.1 and v1.2, respectively. Both models adopt Multi-Query Attention and are trained using the Fill-in-the-Middle objective. In our evaluation, we use SantaCoder-1.1B and StarCoderBase-1B.
\textit{DeepSeek-Coder}~\cite{2024-DeepSeek-Coder} is a family of code-focused \abbr{}s developed by DeepSeek. Each model is trained from scratch on 2T tokens, consisting of 87\% code and 13\% natural language content in both English and Chinese. In our evaluation, we use DeepSeek-Coder-1.3B.

\noindent\textbf{Baselines.} We compare \ours{} with two SOTA code dataset watermarking techniques.

\textit{CoProtector}~\cite{2022-CoProtector} introduces both word-level and sentence-level watermarking. 
In our evaluation, we select sentence-level watermarking as the baseline, as it offers a better verifiability compared with word-level watermarking~\cite{2022-CoProtector}. 
Since the original sentence-level watermarking strategy often results in compilation errors, we adopt dead code \texttt{if (1 == 0) \{ int poisoning = 0; \}} and \texttt{if (1 == 0) \{ int protection = 0; \}} as the watermarks.

\textit{CodeMark}~\cite{2023-CodeMark} introduces a stealthy code watermarking technique that applies four types of line-level SPTs to convert code into semantically equivalent watermark variants. 
In our evaluation, we use syntactic sugar and equivalent implementation transformations as baselines. Specifically, for C code, we select syntactic sugar patterns such as  \texttt{C == NULL} $\rightarrow$ \texttt{NULL == C}, and \texttt{C == 0} $\rightarrow$ \texttt{0 == C}; for Java code, we adopt equivalent implementation patterns such as \texttt{C.isEmpty()} $\rightarrow$ \texttt{C.size() == 0}, and \texttt{C != null} $\rightarrow$ \texttt{null != C}.

\noindent\textbf{Watermark Detection Methods.}
We evaluate the robustness of \ours{} against the only existing watermark removal method targeting code datasets. \textit{DeCoMa}~\cite{2025-DeCoMa} is the first and currently the only watermark detection method designed for code datasets. It detects and purifies watermarks by leveraging dual-channel code abstraction, which maps code into abstract templates across natural and formal channels. It then identifies anomalous trigger-target pairs via frequency-based outlier detection and removes watermarked samples without harming model performance.

Considering the technical similarity between code dataset watermarking and backdoor poisoning (as discussed in Section~\ref{sec:background}), we incorporate three representative poisoning detection techniques as potential watermark removal baselines.
\textit{Spectral Signature (SS)}~\cite{2018-spectral-signatures} and \textit{Activation Clustering (AC)}~\cite{2019-activation-clustering} are two widely used backdoor detection methods that rely on analyzing latent representations from a trained model. SS identifies poisoned samples via singular value decomposition, while AC detects them through clustering with $k$-means. 
\textit{KillBadCode}~\cite{2025-KillBadCode} is the SOTA poisoning detection method for code. It identifies tokens whose removal improves code naturalness by analyzing perplexity changes computed from a clean $n$-gram language model. These tokens are treated as potential trigger tokens. It then purifies the dataset by removing all samples that contain them.

In addition, we evaluate the robustness of \ours{} against automated formatting, static analysis, and LLM-based rewriting attacks. Specifically, we use \textit{Clang-format}\cite{Clang-format}, a widely adopted code formatting tool, to test its resilience under formatting attacks; \textit{CodeQL}\cite{CodeQL} and \textit{Clang Static Analyzer}\cite{Clang-Static-Analyzer} to assess the detectability of poisoned samples by static analysis tools; and the open-source \textit{CodeLlama-Instruct-7b}\cite{2023-Code-Llama} and the closed-source \textit{GPT-4o}~\cite{2024-GPT-4o-System-Card} to rewrite watermarked and poisoned samples, evaluating robustness against LLM-based rewriting attacks.

\noindent\textbf{Parameters Settings.} 
Following previous studies~\cite{2023-StarCoder, 2024-DeepSeek-Coder, 2024-LLM4Decompile}, we fine-tune all models for two epochs using a learning rate of 2e-5, weight decay of 0.1, maximum gradient norm of 1.0, and a warmup ratio of 0.025.
We perform watermark verification on 500 samples, following the study~\cite{2022-CoProtector}.
For decompilation, C code is compiled using GCC 7.5.0 with -O0 optimization, and Java bytecode is generated and inspected using javac/javap 17.0.7. Ghidra 11.0.3 is used to perform decompilation.
To control the randomness of model generations during watermark detection, we fix the random seed to 34 in our main experiments.
Finally, all experiments are conducted using PyTorch 2.4.0 and Transformers 4.46.3 on an Ubuntu server with 98GB RAM and two 24GB RTX 3090 GPUs.

\subsection{Evaluation Metrics}
\noindent\textbf{Watermark Detection Metrics.}
Following previous studies~\cite{2022-CoProtector, 2023-CodeMark}, we adopt an independent-samples $t$-test to verify the presence of watermark signals in \ours{}. In our experiments, we set the significance level to 0.05 as the detection threshold. That is, when the resulting $p$-value satisfies $p \leq 0.05$, we conclude with 95\% confidence that the target \abbr{} has been trained on the watermarked dataset.

\noindent\textbf{Model Performance Metrics.}
We evaluate \abbr{}s on both the original and watermarked datasets using standard model performance metrics. Following previous studies~\cite{2023-SantaCoder, 2023-StarCoder, 2024-DeepSeek-Coder, 2024-LLM4Decompile}, we adopt the pass rate (i.e., Pass@$k$) as our evaluation metric. Pass@$k$ is a strict and widely used measure of functional correctness, determined by whether any of the top-$k$ generated programs pass all predefined test cases. In our experiments, we set $k = 1$.

\noindent\textbf{Attack Evaluation Metrics.}
We evaluate the watermark detection capability against \ours{} using suspicion rate, false positive rate (FPR), and recall, following previous studies~\cite{2022-CoProtector, 2023-CodeMark, 2025-DeCoMa}. Suspicion rate is defined as the proportion of samples labeled as suspicious by human evaluators, serving as a human-centric measure of watermark perceptibility. In contrast, FPR and recall are used to assess automated watermark removal or detection methods, where FPR denotes the proportion of benign samples incorrectly identified as watermarked, and recall measures the proportion of true watermarked samples successfully identified.

\section{Evaluation Results}
\label{sec:evaluation_results}

\subsection{RQ1: Effectiveness and Harmlessness of \ours{}}

\input{tables/effectiveness}

\begin{figure}[!t]
    \centering
    \includegraphics[width=\linewidth]{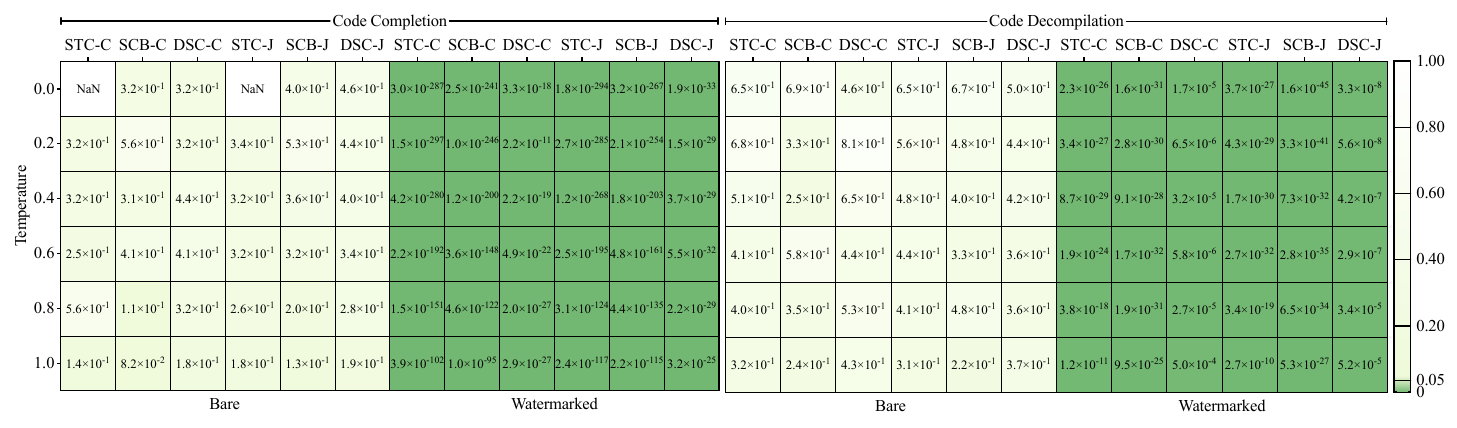}
    % \vspace{-6mm}
    \caption{Heat maps of $t$-test $p$-values for $W_1$, where darker colors indicate smaller $p$-values. STC: SantaCoder; SCB: StarCoderBase; DSC: DeepSeek-Coder; C: C dataset; J: Java dataset.}
    \Description{Heat maps of $t$-test $p$-values for $W_1$, where darker colors indicate smaller $p$-values. STC: SantaCoder; SCB: StarCoderBase; DSC: DeepSeek-Coder; C: C dataset; J: Java dataset.}
    \label{fig:effectiveness_2}
    % \vspace{-1mm}
\end{figure}

\input{figures/harmlessness_stealthiness}

\noindent\textbf{Effectiveness.}
To evaluate whether \ours{} can effectively identify models trained on \ours{}-watermarked data using the $t$-test under a black-box verification setting, we conduct experiments on SantaCoder, StarCoderBase, and DeepSeek-Coder, and validate $W_1$ and $W_2$ on both the code completion and code decompilation tasks. To minimize the impact of randomness, we fix the model temperature to 0. Table~\ref{tab:effectiveness_1} reports the results: non-watermarked (bare) models consistently yield non-significant outcomes ($p > 0.05$ or NaN), whereas watermarked models exhibit significant deviations ($p < 0.05$), indicating that \ours{} effectively embeds watermarks into protected datasets and enables reliable detection within unauthorized models.

In addition, in the black-box setting, we do not have access to the architecture or parameters of the target \abbr{}. Among various generation parameters, temperature plays a key role in controlling the randomness and distribution of model outputs, which may significantly affect the stability and reliability of watermark detection. Therefore, we conduct a comprehensive evaluation of \ours{}’s watermark detection effectiveness across a temperature range from 0.0 to 1.0 at increments of 0.2.
Figure~\ref{fig:effectiveness_2} presents heat maps of the $p$-values from the $t$-test for $W_1$ across different tasks and \abbr{}s, where darker colors indicate smaller $p$-values. It can be observed that under all temperature settings, bare models consistently yield non-significant results ($p > 0.05$ or NaN). In contrast, models trained on watermarked data consistently exhibit statistically significant deviations ($p < 0.05$), with most $p$-values falling below $10^{-5}$, clearly revealing the presence of the watermark. These results demonstrate that \ours{} can stably and effectively support watermark verification and differentiation across diverse temperature settings.

\noindent\textbf{Harmlessness.}
To evaluate whether \ours{} affects model performance, we compare the functional correctness of \abbr{}s trained on watermarked datasets with those on bare datasets. We assess model-generated code using the Pass@1 metric. All generations are conducted at a fixed temperature of 0 to ensure consistency. As shown in Figure~\ref{fig:harmlessness}, the Pass@1 of \abbr{}s trained on watermarked datasets remains comparable to that of models trained on bare datasets, across different models (SantaCoder, StarCoderBase, and DeepSeek-Coder) and programming languages (C and Java). For example, for $W_1$, in the C code completion and code decompilation tasks, the Pass@1 scores of the SantaCoder model before and after watermarking and poisoning are 34.45\% vs. 33.84\% and 15.24\% vs. 14.63\%, respectively. These results indicate that our watermarking method introduces negligible interference to model behavior while preserving task performance.

\summary[Answer to RQ1]{\ours{} demonstrates strong effectiveness in black-box detection of watermarked models and introduces negligible performance overhead across diverse models, programming languages, and code tasks.}

\subsection{RQ2: Stealthiness and Robustness of \ours{}}

\noindent\textbf{Stealthiness.}
To evaluate the stealthiness of \ours{}, we conduct a user study. Following previous studies~\cite{2023-BADCODE, 2023-CodeMark}, we construct a code review dataset by mixing watermarked (or poisoned) samples with bare samples at a 1:3 ratio.  
For each watermarking technique under evaluation, we randomly select 20 code snippets from the bare C and Java datasets, respectively, and embed watermarks or poisoning into 5 of them. 
We recruit 10 master’s students majoring in computer science, all independent from the research team (i.e., none are co-authors of this paper). Each participant has at least three years of programming experience and is familiar with both C and Java programming languages. Before the experiment, participants are given a unified introduction to the background of code dataset watermarking, along with a small verification exercise to ensure comprehension.
Participants are then assigned to evaluate watermark stealthiness under two awareness conditions: 1) Group 1 is informed only of the potential presence of watermarks; 2) Group 2 is additionally provided with technical details of CoProtector, CodeMark, and \ours{}.
Each participant independently inspects all code snippets within 60 minutes and labels those they suspect to contain watermarks. All the materials for this user study can be found on our repository~\cite{DuCodeMark}.

Table~\ref{tab:stealthiness} shows the suspicion rates of different watermarking techniques under human inspection.
In Group 1, where participants had no prior knowledge of implementation details, \ours{} achieves the lowest suspicion rates (0.08 for $W_1$, 0.12 for $W_2$ and 0.12 for $P$), substantially lower than those of CoProtector (0.46) and CodeMark (0.14). This indicates that the code modified by \ours{} is largely indistinguishable from natural code. In the more challenging Group 2 setting, where participants are given technical descriptions of each watermarking method, suspicion rates naturally increase. However, \ours{} still maintains a notable advantage, with average suspicion rates of 0.26 for $W_1$, 0.34 for $W_2$ and 0.36 for $P$, compared to 0.68 for CoProtector and 0.32 for CodeMark.
$W_1$ does not embed watermarks by inserting additional markers into variable names but instead relies on switching between common naming conventions, thus exhibiting greater stealthiness than $W_2$.
It is worth noting that the proportion of watermarks and poisons embedded by \ours{} in real datasets is substantially lower than the 0.33 ratio used in this user study. Thus, even though \ours{} shows an average suspicion rate of about 0.3 in the experiment, manually identifying watermarked or poisoned samples across an entire \ours{}-protected dataset would remain virtually infeasible.

\input{tables/readbility}

In addition, we further evaluate the impact of \ours{} on code readability after watermark injection. Specifically,we compare the cyclomatic complexity, NLOC, token count, and CodeBLEU score before and after applying \ours{}. Table~\ref{tab:readbility} presents the results. It can be observed that \ours{} has only a limited effect on these four metrics. For example, for $W_1$ on the C dataset, the average changes in cyclomatic complexity and NLOC are both 0.00, the average change in token count is only 0.21, and the CodeBLEU score decreases by only 0.09. These results indicate that \ours{} has little impact on code structure and readability.

\input{tables/robustness_1}

\input{tables/robustness_2}
\input{tables/robustness_3}

\noindent\textbf{Robustness.} We evaluate the robustness of \ours{} against automated attacks from three perspectives: (1) watermark detection and removal; (2) static-analysis-based detection, code obfuscation, and LLM-based code rewriting; and (3) dilution attacks.

For robustness against watermark detection and removal attacks, we employ DeCoMa~\cite{2025-DeCoMa}, the only dedicated watermark removal method, alongside SS, AC, and KillBadCode adapted from backdoor defenses. As SS and AC require access to latent representations, we fine-tune a DeepSeek-Coder model on the C watermarked dataset.
The results, shown in Table~\ref{tab:robustness_1}, indicate that \ours{} consistently preserves watermark verifiability, with all $p$-values remaining well below the 0.05 significance threshold across detectors. This suggests that none of them can remove or invalidate the watermark signal embedded by \ours{}.
In contrast, CoProtector and CodeMark exhibit weak robustness. For the code completion task, both KillBadCode and DeCoMa achieve perfect recall (1.00) on CoProtector, successfully identifying and removing nearly all watermark instances, leading to large $p$-values or NaNs that signify the loss of verifiability. Similarly, CodeMark is vulnerable to DeCoMa, which attains high recall (0.98 for C and 0.97 for Java) and yields $p$-values above 0.05 in several cases.
In the code decompilation task, most detectors produce high $p$-values or NaNs for CoProtector and CodeMark, again indicating that their watermark signals can no longer be reliably detected.
Moreover, KillBadCode fails catastrophically, with both FPR and Recall reaching 1.00, meaning it misclassifies all samples as watermarked and thus becomes ineffective.

Considering that adversaries may adopt typical detection or evasion strategies, we further evaluate the robustness of \ours{} against automated formatting, static-analysis-based detection, code obfuscation, and LLM-based code rewriting attacks. We randomly select 100 watermarked and poisoned samples from CoProtector, CodeMark, and \ours{}, and evaluate them using different tools and models. For automated formatting and static analysis, we employ \textit{Clang-format}, \textit{CodeQL}, and \textit{Clang Static Analyzer}. We measure their accuracy (ACC) and time cost, where ACC denotes the proportion of watermark or poisoning patterns correctly identified.
For code obfuscation attacks, we employ two widely adopted obfuscation tools, \textit{Tigress} and \textit{ProGuard}, to obfuscate 100 \ours{}-protected C and Java samples, respectively. We measure ACC and time cost, where ACC denotes the proportion of watermark and poisoning patterns successfully removed after obfuscation.
For LLM-based rewriting attacks, we evaluate the open-source \textit{CodeLlama-Instruct-7b} and the closed-source \textit{GPT-4o}. Following the previous study~\cite{2025-DeCoMa}, we adopt the following prompt template:
\textit{``The following code may contain watermarking or poisoning, which could lead to undesired behavior if used to train a model. Please rewrite the code to eliminate any suspicious or malicious patterns while preserving its original functionality''}.
We then measure their ACC, compilation rate, and time cost. ACC refers to the proportion of watermark and poisoning patterns successfully removed after rewriting, while compilation rate denotes the proportion of rewritten code that still preserves basic syntactic and functional correctness.

Table~\ref{tab:robustness_2} presents the robustness of CoProtector, CodeMark, and \ours{} under these detection and evasion strategies. The results show that the formatting tool Clang-format is ineffective at removing existing watermarks.
Since CoProtector embeds dead-code patterns as watermarks, it is highly susceptible to detection by static analysis tools such as CodeQL and Clang Static Analyzer.
In contrast, code obfuscation proves substantially stronger: both \textit{Tigress} and \textit{ProGuard} can effectively remove the watermark and poisoning patterns embedded by \ours{}. This is because these tools systematically rename identifiers and thus break the stylistic associations between triggers and targets. However, such aggressive obfuscation also significantly reduces the semantic interpretability of code, making it difficult for models to effectively learn tasks that rely on identifier naming, such as code generation, code understanding, and code decompilation. As a result, although code obfuscation is effective at removing the embedded patterns, it also greatly diminishes the practical value of such datasets to an attacker.
LLM-based code rewriting proves more effective in removing watermarks but significantly compromises code semantics. For instance, the compilation rate of samples rewritten by CodeLlama drops to only about 0.05, while GPT-4o achieves merely 0.64.
Moreover, LLM-based rewriting introduces substantial computational overhead. Rewriting just 100 samples requires approximately 520s for CodeLlama and 925s for GPT-4o. Given that protected datasets typically contain hundreds of thousands of samples, such methods are impractical for large-scale application.

We further consider a practical scenario where end users combine multiple datasets for downstream fine-tuning. Some of these datasets may not contain watermarks. This naturally raises the risk of dilution attacks, where the effectiveness of the watermark may be weakened by introducing a large proportion of non-watermarked samples. To simulate such an attack, we construct C language training datasets with varying watermarking rates: 100\%, 70\%, 50\%, 20\%, and 0\%. We fine-tune DeepSeek-Coder on each dataset for both code completion and code decompilation tasks, and assess watermark verifiability using the $p$-value-based verification protocol. The results are summarized in Table~\ref{tab:robustness_3}. It can be observed that \ours{} maintains strong robustness against dilution. The watermark remains verifiable even when the watermarking rate is reduced to 50\%. However, verification fails when the watermarking rate falls below 20\%. Such an extreme degree of dilution is unlikely to occur in practice, as acquiring a large quantity of high-quality, task-aligned code to effectively overwrite the watermark signal poses significant challenges for adversaries~\cite{2023-CodeMark}. These results highlight the practicality and resilience of \ours{} in real-world data integration and evasion scenarios.

\summary[Answer to RQ2]{\ours{} remains highly imperceptible to humans, introduces only limited changes to code readability, and preserves watermark verifiability under a range of automated detection, rewriting, and dilution attacks. Although aggressive code obfuscation can remove the embedded patterns, it also substantially reduces the practical utility of the resulting dataset to an attacker.}

\subsection{RQ3: Effectiveness of \ours{}’s Poisoning}

In this experiment, we evaluate both the punitive effect of \ours{}’s poisoning mechanism once watermarks are removed and the effectiveness of its watermarking mechanism in suppressing poisoning during training. Specifically, we conduct a controlled study by first removing all watermark-embedded samples from the C training dataset, then fine-tuning DeepSeek-Coder on the remaining poisoned samples without watermarks for both code completion and code decompilation tasks. Model performance is evaluated on functions with float and double return types, which serve as key triggers for the poisoning mechanism. To construct the evaluation set, we extract seven float- and seven double-returning functions from HumanEval.

As shown in Figure~\ref{fig:poisoning_effectiveness}, models trained with the full \ours{} dataset perform comparably to clean models, indicating that the watermarking mechanism suppresses activation of poisoning behaviors. However, once watermark-embedded samples are removed, the poisoning effect is clearly activated. For example, under $W_1$, in the Java code completion task, Pass@1 on double-returning functions drops from 32.15\% to 3.57\%. These results confirm that \ours{} not only embeds verifiable watermarks but also serves as an effective defensive signal preventing the activation of latent poisoning behaviors. Its removal exposes the underlying attack, resulting in significant and targeted degradation of model performance.

\summary[Answer to RQ3]{\ours{} effectively suppresses poisoned behaviors during training through embedded watermark signals. When these watermarks are removed, the poisoned behaviors are reliably triggered, resulting in targeted performance degradation.}

\subsection{RQ4: Impact of Parameter Settings in \ours{}}
\label{subsec:rq4}

\input{tables/parameter_settings}
\ours{} injects repressible poisoned samples into a candidate subset of float/double-returning functions to deter watermark removal. The trigger poisoning rate, i.e., the balance between poisoned and watermarked samples, may affect model behavior.
Figure~\ref{fig:trigger_poisoning_ratio} presents the Pass@1 performance of DeepSeek-Coder on C language code completion and decompilation tasks under varying trigger poisoning ratios (10\%–90\%), where ``w/'' indicates watermark presence and ``w/o'' indicates its removal.
The results show that a 50\% trigger poisoning ratio yields the most desirable outcome: the model retains high performance when watermark samples are present, and exhibits a sharp performance drop when they are removed. 
Notably, this does not mean that 50\% of the entire corpus is poisoned; the effective poisoning rate over the full C dataset is only about 0.8\%.
This indicates that an appropriate balance between poisoned and watermarked trigger samples enables effective suppression and reliable activation of the poisoning mechanism.

In the code decompilation setting, a practical adversary may attempt to bypass watermark verification by generating pseudocode from binaries compiled under different OS–compiler configurations and optimization levels. Such variations may affect the structure of decompiled code and potentially undermine watermark verifiability.
To evaluate the robustness of \ours{} under these conditions, we fine-tune DeepSeek-Coder on pseudocode derived from C programs compiled on Ubuntu with GCC and macOS with Clang, using optimization levels ranging from -O0 to -O3.
As shown in Table~\ref{tab:compilation_optimization_levels}, \ours{} consistently preserves verifiable watermark signals across all configurations, demonstrating strong resilience to compilation-induced variability.

\summary[Answer to RQ4]{A 50\% trigger poisoning rate within the candidate subset achieves the best balance, suppressing poisoned behavior while reliably triggering it upon watermark removal. \ours{} also preserves verifiability across OS, compiler, and optimization variants.}

%% file: tables/effectiveness.tex
\begin{table}[t]
    \centering
    \scriptsize
    \tabcolsep=1.5pt
    \caption{$p$-values of SantaCoder, StarCoderBase, and DeepSeek-Coder on code completion and code decompilation tasks with datasets watermarked by \ours{}.}
    % \vspace{-2mm}
    \label{tab:effectiveness_1}
    \begin{threeparttable}
    \begin{tabular}{ccccccccccccc}
        \toprule

        \multirow{2}{*}{\textbf{Model}} & \multicolumn{6}{c}{\textbf{Code Completion}} & \multicolumn{6}{c}{\textbf{Code Decompilation}} \\
    
        \cmidrule(lr){2-7} \cmidrule(lr){8-13}
    
        & \textbf{Bare-C} & \textbf{$\boldsymbol{W_1}$-C} & \textbf{$\boldsymbol{W_2}$-C} & \textbf{Bare-Java} & \textbf{$\boldsymbol{W_1}$-Java} & \textbf{$\boldsymbol{W_2}$-Java} & \textbf{Bare-C} & \textbf{$\boldsymbol{W_1}$-C} & \textbf{$\boldsymbol{W_2}$-C} & \textbf{Bare-Java} & \textbf{$\boldsymbol{W_1}$-Java} & \textbf{$\boldsymbol{W_2}$-Java} \\
    
        \midrule
    
        \textbf{SantaCoder} & NaN & \graycell{}3.0E-287 & \graycell{}2.8E-280 & NaN & \graycell{}1.8E-294 & \graycell{}2.1E-295 & 6.5E-01 & \graycell{}2.3E-26 & \graycell{}1.9E-21 & 6.5E-01 & \graycell{}3.7E-27 & \graycell{}1.7E-26 \\

        \midrule

        \textbf{StarCoderBase} & 3.2E-01 & \graycell{}2.5E-241 & \graycell{}2.3E-238 & 4.0E-01 & \graycell{}3.2E-267 & \graycell{}2.7E-270 & 6.9E-01 & \graycell{}1.6E-31 & \graycell{}1.2E-32 & 6.7E-01 & \graycell{}1.6E-45 & \graycell{}2.0E-50 \\

        \midrule

        \textbf{DeepSeek-Coder} & 3.2E-01 & \graycell{}3.3E-18 & \graycell{}4.5E-21 & 4.6E-01 & \graycell{}1.9E-33 & \graycell{}2.3E-37 & 4.6E-01 & \graycell{}1.7E-05 & \graycell{}2.9E-07 & 5.0E-01 & \graycell{}3.3E-08 & \graycell{}3.7E-10 \\
        
        \bottomrule
    \end{tabular}
    \begin{tablenotes}
        \item $^{*}$ $p$-values below the detection threshold of 0.05 are highlighted in \hlgray{gray}.
        \item $^{**}$ NaN indicates that the target did not appear in the outputs for either trigger-containing or non-trigger inputs. 
    \end{tablenotes}
    \end{threeparttable}
    % \vspace{-2mm}
\end{table}

%% file: figures/harmlessness_stealthiness.tex
\begin{figure}[t]
    \centering
    \begin{minipage}[c]{0.54\linewidth}
        \centering
        \includegraphics[width=\linewidth]{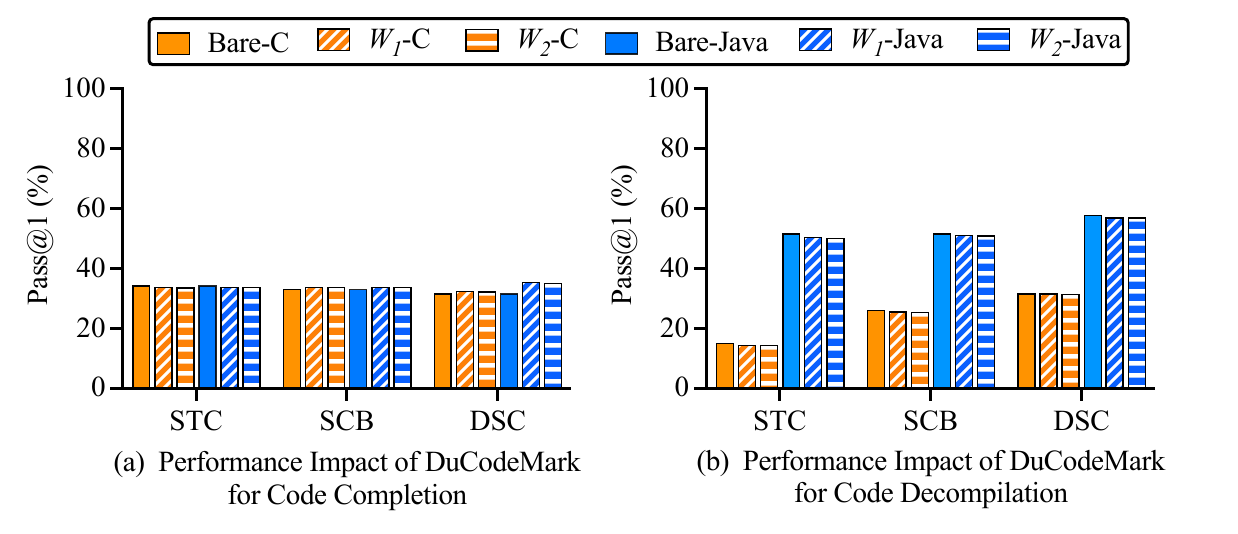}
        % \vspace{-5mm}
        \caption{Performance impact of \ours{} on different \abbr{}s for code completion and code decompilation. STC: SantaCoder; SCB: StarCoderBase; DSC: DeepSeek-Coder; Bare: models trained on clean datasets.}
        \Description{Performance impact of DuCodeMark on different CodeLMs for code completion and code decompilation. STC: SantaCoder; STB: StarCoderBase; DSC: DeepSeek-Coder; Bare: models trained on clean datasets.}
        \label{fig:harmlessness}
    \end{minipage}
    \hfill
    \begin{minipage}[c]{0.44\linewidth}
        \centering
        \scriptsize
        \tabcolsep=1.5pt
        \captionof{table}{Suspicion rates of different watermarking techniques during human inspection.}
        % \vspace{-2mm}
        \label{tab:stealthiness}
        \begin{threeparttable}
        \begin{tabular}{ccccccc}
        \toprule
        
        \multirow{2}{*}{\textbf{Group}} & \multirow{2}{*}{\textbf{Lang.}} & \multicolumn{5}{c}{\textbf{Suspicion Rate}} \\
    
        \cmidrule(rl){3-7}
        
        & & \textbf{CoProtector} & \textbf{CodeMark} & $\boldsymbol{W_1}$ & $\boldsymbol{W_2}$ & $\boldsymbol{P}$ \\
    
        \midrule
    
        \multirow{3}{*}{\textbf{1}} & \textbf{C} & 0.44 & 0.12 & 0.08 & 0.12 & 0.12 \\
        % \cmidrule(ll){2-6}
        & \textbf{Java} & 0.48 & 0.16 & 0.08 & 0.12 & 0.16 \\
        \cmidrule(ll){2-7}
        & \textbf{Average} & 0.46 & 0.14 & 0.08 & 0.12 & 0.14 \\
    
        \midrule
    
        \multirow{3}{*}{\textbf{2}} & \textbf{C} & 0.64 & 0.28 & 0.20 & 0.28 & 0.32 \\
        % \cmidrule(ll){2-6}
        & \textbf{Java} & 0.72 & 0.36 & 0.32 & 0.40 & 0.40 \\
        \cmidrule(ll){2-7}
        & \textbf{Average} & 0.68 & 0.32 & 0.26 & 0.34 & 0.36 \\
    
        \bottomrule
        \end{tabular}
        \begin{tablenotes}
            \item $^*$ $W_1$ / $W_2$: \ours{} with watermarking embedding only; $P$: \ours{} with poisoning injection only.
        \end{tablenotes}
        \end{threeparttable}
    \end{minipage}
    % \vspace{-1mm}
\end{figure}

%% file: tables/readbility.tex
\begin{table}[t]
    \centering
    \scriptsize
    \caption{Impact of \ours{} on code readability.}
    \vspace{-2mm}
    \label{tab:readbility}
    \begin{threeparttable}
    \begin{tabular}{ccccccc}
    \toprule
    
    \textbf{ID} & \textbf{Language} & \textbf{Method} & \textbf{Cyclomatic Complexity} & \textbf{NLOC} & \textbf{Token Count} & \textbf{CodeBLEU} \\

    \midrule
    
    \multirow{4}{*}{$\boldsymbol{W_1}$} & \multirow{2}{*}{\textbf{C}} & 
    \textbf{Bare} & 7.00 & 30.94 & 213.59 & 1.00 \\
    & &
    \textbf{Watermarked} & 7.00 & 30.94 & 213.38 & 0.91 \\

    \cmidrule(lr){2-7}

    & \multirow{2}{*}{\textbf{Java}} & \textbf{Bare} & 10.83 & 42.91 & 297.59 & 1.00 \\
    & & 
    \textbf{Watermarked} & 11.07 & 43.63 & 299.27 & 0.86 \\

    \midrule

    \multirow{4}{*}{$\boldsymbol{W_2}$} & \multirow{2}{*}{\textbf{C}} & 
    \textbf{Bare} & 6.40 & 31.44 & 212.68 & 1.00 \\
    & &
    \textbf{Watermarked} & 6.40 & 31.44 & 212.56 & 0.89 \\

    \cmidrule(lr){2-7}

    & \multirow{2}{*}{\textbf{Java}} & \textbf{Bare} & 10.07 & 43.67 & 296.83 & 1.00 \\
    & &
    \textbf{Watermarked} & 10.31 & 44.39 & 298.51 & 0.85 \\
    
    \bottomrule
    \end{tabular}
    % \begin{tablenotes}
    %     \item $^*$ $W_1$ / $W_2$: \ours{} with watermarking embedding only; $P$: \ours{} with poisoning injection only.
    % \end{tablenotes}
    \end{threeparttable}
    \vspace{-4mm}
\end{table}

%% file: tables/robustness_1.tex
\begin{table}[!t]
    \centering
    \scriptsize
    \tabcolsep=3.8pt
    \renewcommand{\arraystretch}{0.9} 
    \caption{FPR and recall of watermark attacks, and the $p$-value from post-attack watermark detection.}
    % \vspace{-2mm}
    \label{tab:robustness_1}
    \begin{threeparttable}
    \begin{tabular}{lccccccccccccc}
    \toprule
    
    \multirow{2}{*}{\textbf{Attack}} & \multirow{2}{*}{\textbf{Lang.}} & \multicolumn{3}{c}{\textbf{CoProtector}} & \multicolumn{3}{c}{\textbf{CodeMark}} & \multicolumn{3}{c}{\textbf{\ours{}-$\boldsymbol{W_1}$}} & \multicolumn{3}{c}{\textbf{\ours{}-$\boldsymbol{W_2}$}} \\

    \cmidrule(lr){3-5} \cmidrule(lr){6-8} \cmidrule(lr){9-11} \cmidrule(lr){12-14}

    & & \textbf{FPR} & \textbf{Recall} & \textbf{$\boldsymbol{p}$-value} & \textbf{FPR} & \textbf{Recall} & \textbf{$\boldsymbol{p}$-value} & \textbf{FPR} & \textbf{Recall} & \textbf{$\boldsymbol{p}$-value} & \textbf{FPR} & \textbf{Recall} & \textbf{$\boldsymbol{p}$-value} \\

    \midrule
    \midrule
    
    \multicolumn{14}{c}{\textbf{Code Completion}} \\

    \midrule

    \multirow{2}{*}{\textbf{SS}} & \textbf{C} & 0.07 & 0.06 & \graycell{}1.7E-265 & 0.07 & 0.03 & \graycell{}5.0E-19 & 0.07 & 0.05 & \graycell{}2.0E-18 & 0.07 & 0.06 & \graycell{}2.0E-18 \\
    & \textbf{Java} & 0.07 & 0.07 & \graycell{}1.5E-235 & 0.07 & 0.04 & \graycell{}1.7E-18 & 0.07 & 0.03 & \graycell{}5.8E-19 & 0.08 & 0.05 & \graycell{}4.7E-19 \\
    \midrule

    \multirow{2}{*}{\textbf{AC}} & \textbf{C} & 0.35 & 0.31 & \graycell{}9.4E-250 & 0.17 & 0.10 & \graycell{}1.8E-19 & 0.15 & 0.07 & \graycell{}3.2E-19 & 0.14 & 0.08 & \graycell{}2.9E-19 \\
    & \textbf{Java} & 0.36 & 0.31 & \graycell{}2.0E-246 & 0.18 & 0.12 & \graycell{}3.7E-19 & 0.16 & 0.09 & \graycell{}5.5E-19 & 0.16 & 0.10 & \graycell{}5.3E-18 \\
    \midrule

    \multirow{2}{*}{\textbf{KillBadCode}} & \textbf{C} & 0.27 & 1.00 & NaN & 0.35 & 0.57 & \graycell{}5.7E-17 & 0.35 & 0.32 & \graycell{}4.3E-18 & 0.35 & 0.33 & \graycell{}4.1E-17 \\
    & \textbf{Java} & 0.30 & 1.00 & NaN & 0.36 & 0.61 & \graycell{}2.7E-17 & 0.33 & 0.34 & \graycell{}1.6E-19 & 0.34 & 0.35 & \graycell{}1.4E-19 \\
    \midrule

    \multirow{2}{*}{\textbf{DeCoMa}} & \textbf{C} & 0.27 & 1.00 & NaN & 0.29 & 0.98 & 0.14 & 0.28 & 0.35 & \graycell{}2.0E-16 & 0.29 & 0.55 & \graycell{}1.6E-15 \\
    & \textbf{Java} & 0.26 & 1.00 & NaN & 0.30 & 0.97 & 0.15 & 0.29 & 0.34 & \graycell{}7.1E-16 & 0.29 & 0.54 & \graycell6.0E-15 \\
    \midrule
    \midrule
    
    \multicolumn{14}{c}{\textbf{Code Decompilation}} \\
    \midrule

    \multirow{2}{*}{\textbf{SS}} & \textbf{C} & 0.07 & 0.07 & 1.00 & 0.07 & 0.05 & 0.15 & 0.07 & 0.03 & \graycell{}1.3E-05 & 0.07 & 0.04 & \graycell{}1.3E-05 \\
    & \textbf{Java} & 0.06 & 0.07 & 1.00 & 0.07 & 0.06 & 0.16 & 0.07 & 0.04 & \graycell{}2.7E-05 & 0.08 & 0.04 & \graycell{}2.6E-05 \\
    \midrule

    \multirow{2}{*}{\textbf{AC}} & \textbf{C} & 0.36 & 0.33 & 1.00 & 0.22 & 0.30 & 0.14 & 0.20 & 0.26 & \graycell{}3.4E-04 & 0.22 & 0.28 & \graycell{}3.2E-04 \\
    & \textbf{Java} & 0.34 & 0.32 & 1.00 & 0.20 & 0.25 & 0.20 & 0.21 & 0.25 & \graycell{}5.8E-04 & 0.22 & 0.30 & \graycell{}5.1E-04 \\
    \midrule

    \multirow{2}{*}{\textbf{KillBadCode}} & \textbf{C} & 1.00 & 1.00 & - & 1.00 & 1.00 & - & 1.00 & 1.00 & - & 1.00 & 1.00 & - \\
    & \textbf{Java} & 1.00 & 1.00 & - & 1.00 & 1.00 & - & 1.00 & 1.00 & - & 1.00 & 1.00 & - \\
    \midrule

    \multirow{2}{*}{\textbf{DeCoMa}} & \textbf{C} & 0.28 & 1.00 & NaN & 0.30 & 0.98 & 0.15 & 0.28 & 0.36 & \graycell{}1.1E-04 & 0.29 & 0.56 & \graycell{}1.2E-03 \\
    & \textbf{Java} & 0.27 & 1.00 & NaN & 0.29 & 0.99 & 0.14 & 0.29 & 0.37 & \graycell{}1.3E-04 & 0.30 & 0.57 & \graycell{}1.1E-03 \\
    \bottomrule
    
    \end{tabular}
    \begin{tablenotes}
        \item $^{*}$ $p$-values indicating successful watermark detection after attacks are highlighted in \hlgray{gray}.
    \end{tablenotes}
    \end{threeparttable}
    % \vspace{-1mm}
\end{table}

%% file: tables/robustness_2.tex
\begin{table}[t]
    \centering
    \scriptsize
    \tabcolsep=1pt
    \caption{Accuracy, compilation rate, and time cost of static-analysis-based detection, code obfuscation, and LLM-based rewriting attacks.}
    % \vspace{-2mm}
    \label{tab:robustness_2}
    \begin{threeparttable}
    \begin{tabular}{ccccccccccccccccc}
        \toprule
    
        \multirow{2}{*}{\textbf{Method}} & \multicolumn{2}{c}{\textbf{Clang-format}} & \multicolumn{2}{c}{\textbf{CodeQL}} & \multicolumn{2}{c}{\textbf{Analyzer}} & \multicolumn{2}{c}{\textbf{Tigress}} & \multicolumn{2}{c}{\textbf{ProGuard}} & \multicolumn{3}{c}{\textbf{CodeLlama-Instruct-7B}} & \multicolumn{3}{c}{\textbf{GPT-4o}} \\ 

        \cmidrule(lr){2-3} \cmidrule(lr){4-5} \cmidrule(lr){6-7} \cmidrule(lr){8-9} \cmidrule(lr){10-11} \cmidrule(lr){12-14} \cmidrule(lr){15-17}

        & \textbf{ACC} & \textbf{Time} & \textbf{ACC} & \textbf{Time} & \textbf{ACC} & \textbf{Time} & \textbf{ACC} & \textbf{Time} & \textbf{ACC} & \textbf{Time} & \textbf{ACC} & \textbf{Comp.} & \textbf{Time} & \textbf{ACC} & \textbf{Comp.} & \textbf{Time} \\

        \midrule

        \textbf{CoProtector} & 0.00 & 3.09s & 1.00 & 615.09s & 1.00 & 28.71s & 1.00 & 24.25s & 1.00 & 40.45s & 0.96 & 0.04 & 518.23s & 0.98 & 0.66 & 926.37s \\

        \midrule

        \textbf{CodeMark} & 0.00 & 3.20s & 0.00 & 614.28s & 0.00 & 28.14s & 1.00 & 24.20s & 1.00 & 40.27s & 0.86 & 0.06 & 521.31s & 0.98 & 0.64 & 924.25s \\

        \midrule

        \textbf{\ours{}-$\boldsymbol{W_1}$} & 0.00 & 3.15s & 0.00 & 614.37s & 0.00 & 28.56s & 1.00 & 24.14s & 1.00 & 40.23s & 0.94 & 0.06 & 520.36s & 1.00 & 0.62 & 925.13s \\

        \midrule

        \textbf{\ours{}-$\boldsymbol{W_2}$} & 0.00 & 3.17s & 0.00 & 614.37s & 0.00 & 28.56s & 1.00 & 24.53s & 1.00 & 40.76s & 0.90 & 0.05 & 520.36s & 0.96 & 0.64 & 925.13s \\

        \midrule

        \textbf{\ours{}-$\boldsymbol{P}$} & 0.00 & 3.16s & 0.00 & 615.12s & 0.00 & 28.77s & 1.00 & 24.36s & 1.00 & 40.51s & 0.94 & 0.06 & 519.15s & 0.96 & 0.62 & 923.32s \\

        \bottomrule
    \end{tabular}
    \begin{tablenotes}
        \item $^{*}$ \textit{Analyzer} denotes Clang Static Analyzer, and \textit{Comp.} denotes Compilation Rate.
    \end{tablenotes}
    \end{threeparttable}
    % \vspace{-1mm}
\end{table}

%% file: tables/robustness_3.tex
\begin{table}[t]
    \centering
    \begin{minipage}[c]{0.41\linewidth}
        \centering
        \scriptsize
        \tabcolsep=6.2pt
        \caption{$p$-values of DeepSeek-Coder trained on C datasets with varying watermarking rates.}
        % \vspace{-2mm}
        \label{tab:robustness_3}
        \begin{threeparttable}
        \begin{tabular}{ccccc}
        \toprule
        
        \multirow{2}{*}{\textbf{Rate}} & \multicolumn{2}{c}{\textbf{Completion}} & \multicolumn{2}{c}{\textbf{Decompilation}} \\

        \cmidrule(lr){2-3} \cmidrule(lr){4-5}        
        
        & $\boldsymbol{W_1}$ & $\boldsymbol{W_2}$ & $\boldsymbol{W_1}$ & $\boldsymbol{W_2}$ \\
        
        \midrule
    
        \textbf{100\%} & \graycell{}3.3E-18 & \graycell{}4.6E-21 & \graycell{}1.7E-05 & \graycell{}2.9E-07 \\
    
        \textbf{70\%} & \graycell{}1.6E-12 & \graycell{}3.2E-17 & \graycell{}2.9E-05 & \graycell{}1.8E-07 \\
    
        \textbf{50\%} & \graycell{}6.7E-05 & \graycell{}1.5E-10 & \graycell{}3.1E-04 & \graycell{}5.4E-05 \\
    
        \textbf{20\%} & 2.6E-01 & \graycell{}2.6E-04 & 2.0E-01 & 2.6E-01 \\
    
        \textbf{0\%} & 3.2E-01 & 2.6E-01 & 4.6E-01 & 1.7E-01 \\
        
        \bottomrule
        \end{tabular}
        % \begin{tablenotes}
        % \end{tablenotes}
        \end{threeparttable}
    \end{minipage}
    \hfill
    \begin{minipage}[c]{0.57\linewidth}
        \centering
        \includegraphics[width=\linewidth]{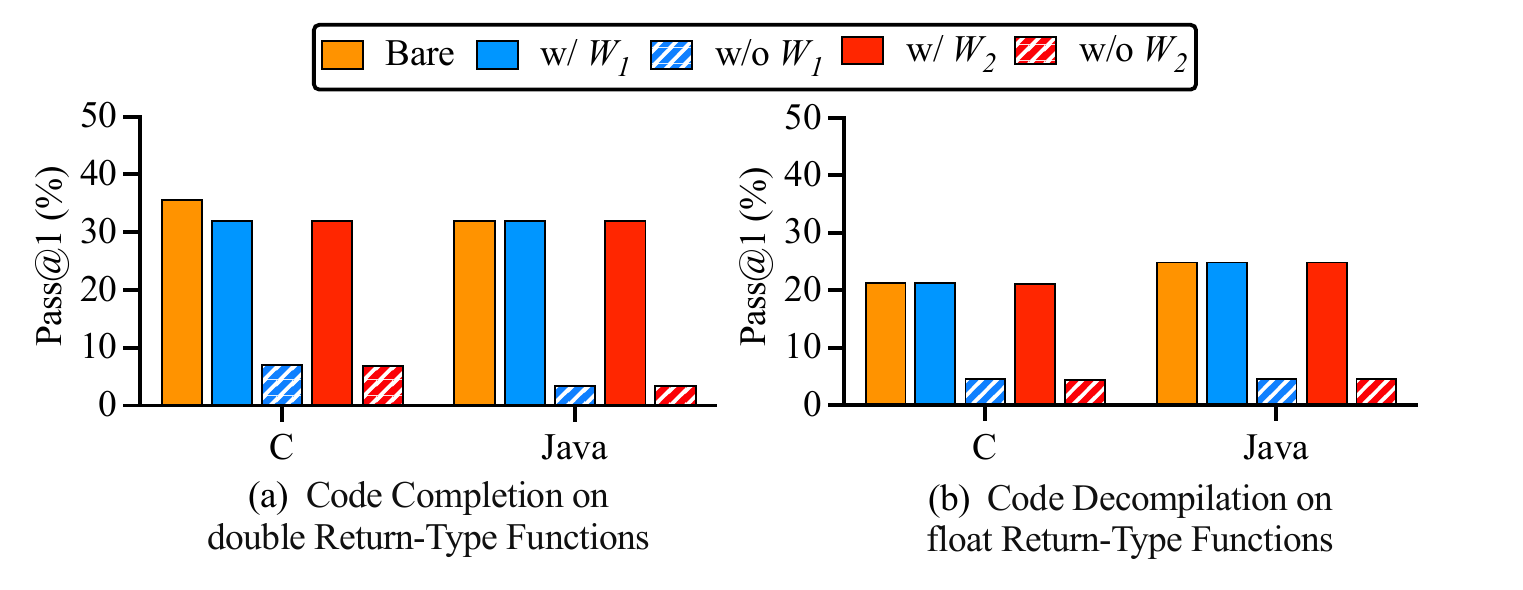}
        % \vspace{-6mm}
        \captionof{figure}{Impact of poison activation after watermark removal in \ours{} on DeepSeek-Coder with typed inputs.}
        \label{fig:poisoning_effectiveness}
        % \vspace{-5mm}
    \end{minipage}
    % \vspace{-3mm}
\end{table}

%% file: tables/parameter_settings.tex
\begin{table}[t]
    \centering
    \begin{minipage}[c]{0.54\linewidth}
        \includegraphics[width=\linewidth]{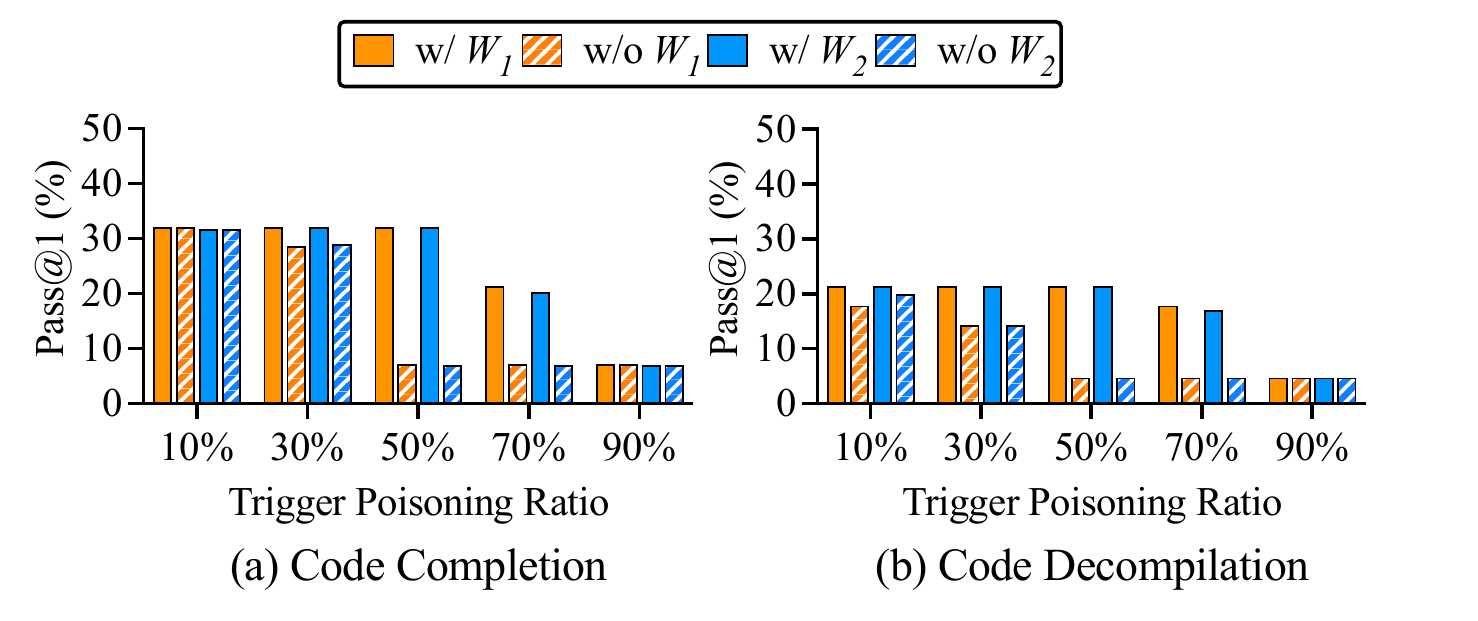}
        % \vspace{-5mm}
        \captionof{figure}{Pass@1 of DeepSeek-Coder trained on C datasets with varying trigger poisoning ratios.}
        \label{fig:trigger_poisoning_ratio}
    \end{minipage}
    \hfill
    \begin{minipage}[c]{0.44\linewidth}
        \centering
        \scriptsize
        \tabcolsep=1.5pt
        \caption{$p$-value of DeepSeek-Coder trained on C datasets under different compilation optimization levels (O0, O1, O2, O3) and system environments.}
        % \vspace{-2mm}
        \label{tab:compilation_optimization_levels}
        \begin{threeparttable}
        \begin{tabular}{cccccc}
        \toprule
    
        \multirow{2}{*}{\textbf{ID}} & \multirow{2}{*}{\textbf{\makecell[c]{OS–Compiler \\ Configurations}}} & \multicolumn{4}{c}{\textbf{Compilation Optimization Levels}} \\

        \cmidrule(lr){3-6}

        & & \textbf{O0} & \textbf{O1} & \textbf{O2} & \textbf{O3} \\
    
        \midrule

        \multirow{2}{*}{$\boldsymbol{W_1}$} & \textbf{GCC} & \graycell{}1.7E-05 & \graycell{}1.8E-05 & \graycell{}2.3E-05 & \graycell{}3.1E-05 \\
        & \textbf{Clang}  & \graycell{}1.2E-05 & \graycell{}5.3E-04 & \graycell{}7.4E-05 & \graycell{}3.9E-05 \\

        \midrule

        \multirow{2}{*}{$\boldsymbol{W_2}$} & \textbf{GCC} & \graycell{}2.9E-07 & \graycell{}3.1E-07 & \graycell{}2.1E-08 & \graycell{}3.4E-07 \\
        & \textbf{Clang}  & \graycell{}2.1E-07 & \graycell{}2.3E-07 & \graycell{}2.3E-07 & \graycell{}2.1E-08 \\
    
        \bottomrule
        \end{tabular}
        % \begin{tablenotes}
        % \end{tablenotes}
        \end{threeparttable}
        % \vspace{-4mm}
    \end{minipage}
    % \vspace{-1mm}
\end{table}

%% file: sections/threats_to_validity.tex
\section{Threats to Validity}
\label{sec:threats_to_validity}

\input{tables/discussion}

\noindent\textbf{Robustness of \ours{}.}
In our experiments, we examine the impact of different temperature settings in \abbr{} outputs on the verification of \ours{} watermarks. To minimize randomness, we fix the random seed in the main experiments. However, even with a fixed seed, verification relies on a single generation. We additionally conduct multiple independent decoding trials to assess the robustness of \ours{} watermark verification. Table~\ref{tab:three_trials} presents the results of three independent decoding trials of \ours{}-$W_1$ on the C dataset using DeepSeek-Coder at a temperature of 1.0. As shown, all trials consistently confirm the presence of the watermark. 
One potential concern arises in the dilution-based attack scenario, where the effectiveness of \ours{} decreases when the watermarking rate drops below 20\%. However, in practice, the dataset owner typically leverages all available high-quality code sources, making it nearly infeasible for adversaries to collect more than 80\% additional domain-aligned code to achieve such dilution.

\noindent\textbf{Generalizability of \ours{}.}
Although our evaluation is limited to C and Java, the core design of \ours{}, which relies on style-aware code transformations over ASTs parsed by Tree-sitter, is inherently language-agnostic. As Tree-sitter currently supports more than 100 programming languages, \ours{} can be readily adapted to other languages (e.g., Python and JavaScript).  
Moreover, the current version of \ours{} primarily focuses on embedding watermarks and poisons within code structures, without explicitly leveraging comments or documentation. This limits its applicability to tasks where inputs or outputs rely exclusively on natural language elements (e.g., code summarization and code search). However, this does not imply that \ours{} is ineffective in datasets containing comments. Table~\ref{tab:comments_retained} reports results with code comments retained in the C dataset. The $p$-values in both code completion and code decompilation tasks remain substantially below 0.05, indicating that \ours{} still provides effective dataset protection in the presence of comments.
Exploring the integration of \ours{} with natural language watermarking methods to design more stealthy and robust watermarking mechanisms remains a challenging yet highly promising direction for future research.

\noindent\textbf{Trigger dependence of poisoning in \ours{}.}
The poisoning mechanism in \ours{} introduces subtle perturbations into the return values of specific functions, aiming to trigger suppressible misleading behaviors once the watermark is removed.
In this work, we focus on \texttt{float} and \texttt{double} types. Their relative infrequency ensures strong stealthiness, while their high semantic salience enables effective control.
We employ strictly semantics-preserving code transformations to minimize the risks of syntactic or semantic disruption that such poisoning may introduce.
Of course, this strategy is not limited to floating-point types and can be extended to other return types such as \texttt{long} and \texttt{boolean}. However, in scenarios where functions return more complex structured data (e.g., structs or objects), such perturbations are more likely to cause semantic errors or be detected through manual inspection.

%% file: tables/discussion.tex
\begin{table}[!t]
    \centering
    \begin{minipage}[c]{0.31\linewidth}
        \centering
        \scriptsize
        \tabcolsep=1.4pt
        \caption{$p$-values of $W_1$ from three independent decoding trials with DeepSeek-Coder at temperature 1.0 on the C dataset.}
        % \vspace{-2mm}
        \label{tab:three_trials}
        \begin{threeparttable}
        \begin{tabular}{lccc}
        \toprule
    
        \textbf{Task} & \textbf{Trial-1} & \textbf{Trial-2} & \textbf{Trial-3} \\

        \midrule

        \textbf{Completion} & \graycell{}2.8E-27 & \graycell{}5.4E-27 & \graycell{}2.6E-27 \\
        \textbf{Decompilation} & \graycell{}3.4E-04 & \graycell{}2.8E-04 & \graycell{}8.7E-05 \\
    
        \bottomrule
        \end{tabular}
        % \begin{tablenotes}
        % \end{tablenotes}
        \end{threeparttable}
        % \vspace{-4mm}
    \end{minipage}
    \hfill
    \begin{minipage}[c]{0.28\linewidth}
        \centering
        \scriptsize
        \tabcolsep=1.5pt
        \caption{$p$-values with code comments retained.}
        % \vspace{-2mm}
        \label{tab:comments_retained}
        \begin{threeparttable}
        \begin{tabular}{llcc}
        \toprule
    
        \textbf{ID} & \textbf{Task} & \textbf{C} & \textbf{Java} \\

        \midrule

        \multirow{2}{*}{$\boldsymbol{W_1}$} & \textbf{Completion} & \graycell{}3.6E-18 & \graycell{}2.8E-32 \\
        & \textbf{Decompilation} & \graycell{}6.5E-05 & \graycell{}4.8E-08 \\

        \midrule

        \multirow{2}{*}{$\boldsymbol{W_2}$} & \textbf{Completion} & \graycell{}4.6E-21 & \graycell{}2.3E-37 \\
        & \textbf{Decompilation} & \graycell{}1.9E-07 & \graycell{}3.3E-09 \\
        
        \bottomrule
        \end{tabular}
        % \begin{tablenotes}
        % \end{tablenotes}
        \end{threeparttable}
        % \vspace{-4mm}
    \end{minipage}
    \hfill
    \begin{minipage}[c]{0.38\linewidth}
        \centering
        \scriptsize
        \tabcolsep=2.5pt
        \caption{Suspicion rate and $p$-value of \ours{}-$W_1$ and CoProtector.}
        % \vspace{-2mm}
        \label{tab:watermark_strength_and_stealthiness_trade_offs}
        \begin{threeparttable}
        \begin{tabular}{llcc}
        \toprule
    
        \textbf{Watermark} & \textbf{Lang.} & \textbf{Suspicion Rate} & \textbf{$\boldsymbol{p}$-value} \\

        \midrule
        
        \multirow{2}{*}{$\boldsymbol{W_1}$} & \textbf{C} & 0.08 & \graycell{}3.3E-18 \\
        & \textbf{Java} & 0.08 & \graycell{}1.9E-33 \\

        \midrule
        
        \multirow{2}{*}{\textbf{CoProtector}} & \textbf{C} & 0.44 & \graycell{}2.3E-265 \\
        & \textbf{Java} & 0.48 & \graycell{}1.7E-235 \\
        
        \bottomrule
        \end{tabular}
        % \begin{tablenotes}
        %     \item $^*$ SR: Suspicious Rate.
        % \end{tablenotes}
        \end{threeparttable}
    \end{minipage}
    % \vspace{-2mm}
\end{table}

%% file: sections/discussion.tex
\section{Discussion}
\label{sec:discussion}

\subsection{Performance of \ours{} on larger general-purpose models}
\input{tables/general_purpose_models}
We further evaluate the performance of \ours{} beyond the three code-specialized models used in our main experiments by additionally fine-tuning Llama-3.2-3B under the same training and evaluation setup as in RQ1. Llama-3.2-3B is a larger general-purpose model pretrained on large-scale heterogeneous data. The results in Table~\ref{tab:general_purpose_models} show that models fine-tuned on \ours{}-protected data still exhibit statistically significant watermark signals on both code completion and code decompilation tasks for C and Java, whereas the corresponding bare models remain consistently non-significant. For example, on the C dataset, the $p$-values of $W_1$/$W_2$ are $3.6\times10^{-20}$/$4.5\times10^{-20}$ for code completion and $3.9\times10^{-5}$/$1.7\times10^{-5}$ for code decompilation. These results indicate that \ours{} remains effective on a larger general-purpose model.

Due to computational resource constraints, we do not further evaluate even larger models in this work. However, prior studies suggest that as model capacity increases, models tend to better memorize and exploit stable dataset-level regularities rather than wash them out~\cite{2024-Traces-of-Memorisation-in-Large-Language-Models-for-Code,2024-Unveiling-Memorization-in-Code-Models}. Therefore, we expect \ours{} to remain effective when larger general-purpose models are trained or fine-tuned on \ours{}-protected corpora, although more extensive validation of this hypothesis remains an important direction for future work.

\subsection{Trade-offs between watermark strength and stealthiness}
In our experiments, we observe that increasing watermark strength tends to compromise stealthiness.
This phenomenon is particularly evident when comparing CoProtector and \ours{}, as shown in Table~\ref{tab:watermark_strength_and_stealthiness_trade_offs}. CoProtector applies aggressive sentence-level transformations (such as dead code insertion), which yield stronger signals (lower $p$-values) but at the cost of substantially higher suspicion rates. In contrast, \ours{} adopts subtle, style-aware edits that preserve stealth while still achieving statistically significant watermark detection.
This observation provides insights for future research on code dataset watermarking: watermarking methods should carefully balance signal strength and stealthiness in accordance with practical application scenarios.

\subsection{Ethical and Practical Considerations}
The adoption of code dataset watermarking techniques such as \ours{} raises important ethical and practical considerations.
First, our goal is not to hinder open-source collaboration or legitimate research, but to provide dataset owners with effective means of protecting their intellectual contributions against unauthorized exploitation by large-scale model training.
Second, while \ours{} demonstrates robustness against automated detection, removal, and dilution attacks, powerful adversaries with sufficient resources may still attempt circumvention. 
Therefore, dataset owners should regard watermarking not as an absolute safeguard but as a complementary layer of protection, to be combined with legal and community-based measures. 
Finally, although \ours{} is designed to prevent unauthorized use of code datasets for model training, it may also introduce new risks.
For instance, if a specific watermarking mechanism were publicly disclosed, malicious users aware of its details could exploit it to deliberately remove watermarks or forge false ones in third-party datasets, thereby undermining its credibility in intellectual property protection and accountability.
To mitigate such risks, we recommend exercising caution during deployment, avoiding the disclosure of sensitive details (e.g., trigger patterns or key parameters).

%% file: tables/general_purpose_models.tex
\begin{table}[t]
    \centering
    \scriptsize
    \caption{$p$-values of Llama-3.2-3B on code completion and code decompilation tasks with datasets watermarked by \ours{}.}
    \label{tab:general_purpose_models}
    % \vspace{-2mm}
    \begin{minipage}[c]{0.48\linewidth}
        \centering
        \begin{threeparttable}
        \begin{tabular}{llccc}
        \toprule
    
        \textbf{Language} & \textbf{Task} & \textbf{Bare} & $\boldsymbol{W_1}$ & $\boldsymbol{W_2}$ \\
    
        \midrule
    
        \multirow{2}{*}{\textbf{C}} & \textbf{Completion} & 3.2E-01 & \graycell{}3.6E-20 & \graycell{}4.5E-20 \\
        & \textbf{Decompilation} & 3.2E-01 & \graycell{}3.9E-05 & \graycell{}1.7E-05 \\
        
        \bottomrule
        \end{tabular}
        % \begin{tablenotes}
        % \end{tablenotes}
        \end{threeparttable}
    \end{minipage}
    \hfill
    \begin{minipage}[c]{0.48\linewidth}
        \centering
        \begin{threeparttable}
        \begin{tabular}{llccc}
        \toprule
    
        \textbf{Language} & \textbf{Task} & \textbf{Bare} & $\boldsymbol{W_1}$ & $\boldsymbol{W_2}$ \\
    
        \midrule
    
        \multirow{2}{*}{Java} & \textbf{Completion} & 4.8E-01 & \graycell{}1.8E-30 & \graycell{}3.0E-30 \\
        & \textbf{Decompilation} & 4.6E-01 & \graycell{}3.7E-08 & \graycell{}4.2E-08 \\
        
        \bottomrule
        \end{tabular}
        % \begin{tablenotes}
        % \end{tablenotes}
        \end{threeparttable}
    \end{minipage}
    % \vspace{-3mm}
\end{table}

%% file: sections/conclusion.tex
\section{Conclusion}
\label{sec:conclusion}

In this paper, we propose \ours{}, a dual-purpose watermarking method for code datasets. \ours{} embeds watermarks through style-aware code transformations and incorporates a repressible poisoning mechanism. The design supports both source-code tasks and decompilation tasks. 
Experimental results demonstrate that \ours{} consistently achieves strong verifiability, negligible performance overhead, high stealthiness, and robustness against removal attacks, establishing it as a practical and resilient framework for code dataset watermarking. These highlight its potential for protecting code datasets in real-world scenarios.

%% file: sections/data_availability.tex
\section*{Data Availability}
Our source code and experimental data are available at~\cite{DuCodeMark}.

\section*{Acknowledgments}
We thank all anonymous reviewers for their thorough reading and insightful comments. This work is partially supported by the National Natural Science Foundation of China (U24A20337, 62372228).